\pdfoutput=1
\documentclass[12pt,a4paper]{article}

\usepackage{ifthen} 
\newboolean{pdflatex}
\setboolean{pdflatex}{true} 

\newboolean{articletitles}
\setboolean{articletitles}{true} 

\newboolean{uprightparticles}
\setboolean{uprightparticles}{false} 

\def\paperauthors{LHCb collaboration} 
\def\paperasciititle{Measurement of the W boson cross-section as a function of the muon transverse momentum in proton-proton collisions at 5.02 TeV} 
\def\papertitle{Measurement of the $W \to \mu\nu$ cross-sections as a function of the muon transverse momentum in $pp$ collisions at 5.02 TeV } 
\def\paperkeywords{{High Energy Physics}, {LHCb}} 
\def\papercopyright{\the\year\ CERN for the benefit of the LHCb collaboration} 
\def\paperlicence{CC BY 4.0 licence}
\def\paperlicenceurl{https://creativecommons.org/licenses/by/4.0/}

\newif\ifEnableSectionTOCLinks
\EnableSectionTOCLinksfalse 

\usepackage[top=1in, bottom=1.25in, left=1in, right=1in]{geometry}

\usepackage{microtype}
\usepackage{lineno}  
\usepackage{xspace} 
\usepackage{caption} 

\usepackage{graphicx}  
\usepackage{color}
\usepackage{colortbl}
\graphicspath{{./figs/}} 

\usepackage{amsmath} 
\usepackage{amssymb}
\usepackage{amsfonts}
\usepackage{upgreek} 

\newcommand*\patchAmsMathEnvironmentForLineno[1]{%
\expandafter\let\csname old#1\expandafter\endcsname\csname #1\endcsname
\expandafter\let\csname oldend#1\expandafter\endcsname\csname
end#1\endcsname
 \renewenvironment{#1}%
   {\linenomath\csname old#1\endcsname}%
   {\csname oldend#1\endcsname\endlinenomath}%
}
\newcommand*\patchBothAmsMathEnvironmentsForLineno[1]{%
  \patchAmsMathEnvironmentForLineno{#1}%
  \patchAmsMathEnvironmentForLineno{#1*}%
}
\AtBeginDocument{%
\patchBothAmsMathEnvironmentsForLineno{equation}%
\patchBothAmsMathEnvironmentsForLineno{align}%
\patchBothAmsMathEnvironmentsForLineno{flalign}%
\patchBothAmsMathEnvironmentsForLineno{alignat}%
\patchBothAmsMathEnvironmentsForLineno{gather}%
\patchBothAmsMathEnvironmentsForLineno{multline}%
\patchBothAmsMathEnvironmentsForLineno{eqnarray}%
}

\usepackage[pdftex,
            pdfauthor={\paperauthors},
            pdftitle={\paperasciititle},
            pdfkeywords={\paperkeywords}]{hyperref}
\usepackage{hyperxmp}
\hypersetup{
    pdfcopyright={Copyright (C) \papercopyright},
    pdflicenseurl={\paperlicenceurl}
}

\usepackage[bottom,flushmargin,hang,multiple]{footmisc}

\usepackage[all]{hypcap} 

\usepackage{xspace}
\usepackage{upgreek}

\def\lhcb   {\mbox{LHCb}\xspace}

\def\MagUp {\mbox{\em Mag\kern -0.05em Up}\xspace}

\ifthenelse{\boolean{uprightparticles}}%
{

 \def\Pmu         {\ensuremath{\upmu}\xspace}
 \def\Pnu         {\ensuremath{\upnu}\xspace}

 \def\Ptau        {\ensuremath{\uptau}\xspace}

 \def\PDelta      {\ensuremath{\Delta}\xspace}
 \def\PXi         {\ensuremath{\Xi}\xspace}
 \def\PLambda     {\ensuremath{\Lambda}\xspace}
 \def\PSigma      {\ensuremath{\Sigma}\xspace}
 \def\POmega      {\ensuremath{\Omega}\xspace}
 \def\PUpsilon    {\ensuremath{\Upsilon}\xspace}
 \let\oldPi\Pi
 \def\PPi         {\ensuremath{\oldPi}\xspace}

 \def\PB      {\ensuremath{\mathrm{B}}\xspace}
 \def\PD      {\ensuremath{\mathrm{D}}\xspace}

 \def\PK      {\ensuremath{\mathrm{K}}\xspace}
 \def\PW      {\ensuremath{\mathrm{W}}\xspace}
 \def\PZ      {\ensuremath{\mathrm{Z}}\xspace}
 \def\Ps      {\ensuremath{\mathrm{s}}\xspace}

 \def\thebaroffset{0.0em}
}
{

 \def\Pmu         {\ensuremath{\mu}\xspace}
 \def\Pnu         {\ensuremath{\nu}\xspace}

 \def\Ptau        {\ensuremath{\tau}\xspace}

 \mathchardef\PDelta="7101
 \mathchardef\PXi="7104
 \mathchardef\PLambda="7103
 \mathchardef\PSigma="7106
 \mathchardef\POmega="710A
 \mathchardef\PUpsilon="7107
 \mathchardef\PPi="7105
 \def\PB      {\ensuremath{B}\xspace}
 \def\PD      {\ensuremath{D}\xspace}

 \def\PK      {\ensuremath{K}\xspace}
 \def\PW      {\ensuremath{W}\xspace}
 \def\PZ      {\ensuremath{Z}\xspace}
 \def\Ps      {\ensuremath{s}\xspace}

 \def\thebaroffset{0.18em}
}
\newcommand{\offsetoverline}[2][\thebaroffset]{\kern #1\overline{\kern -#1 #2}}%

\makeatletter
\ifcase \@ptsize \relax
  \newcommand{\miniscule}{\@setfontsize\miniscule{4}{5}}
\or
  \newcommand{\miniscule}{\@setfontsize\miniscule{5}{6}}
\or
  \newcommand{\miniscule}{\@setfontsize\miniscule{5}{6}}
\fi
\makeatother

\DeclareRobustCommand{\optbar}[1]{\shortstack{{\miniscule (\rule[.5ex]{1.25em}{.18mm})}
  \\ [-.7ex] $#1$}}

\def\muon       {{\ensuremath{\Pmu}}\xspace}
\def\mup        {{\ensuremath{\Pmu^+}}\xspace}
\def\mun        {{\ensuremath{\Pmu^-}}\xspace} 

\def\mumu       {{\ensuremath{\Pmu^+\Pmu^-}}\xspace}

\def\tautau     {{\ensuremath{\Ptau^+\Ptau^-}}\xspace}

\def\neu        {{\ensuremath{\Pnu}}\xspace}
\def\neub       {{\ensuremath{\overline{\Pnu}}}\xspace}
\def\neum       {{\ensuremath{\neu_\mu}}\xspace}
\def\neumb      {{\ensuremath{\neub_\mu}}\xspace}

\def\W      {{\ensuremath{\PW}}\xspace}
\def\Wp     {{\ensuremath{\PW^+}}\xspace}
\def\Wm     {{\ensuremath{\PW^-}}\xspace}

\def\Z      {{\ensuremath{\PZ}}\xspace}

\def\squark    {{\ensuremath{\Ps}}\xspace}

\def\KorKbar {\kern \thebaroffset\optbar{\kern -\thebaroffset \PK}{}\xspace}

\def\D       {{\ensuremath{\PD}}\xspace}

\def\DorDbar {\kern \thebaroffset\optbar{\kern -\thebaroffset \PD}\xspace}

\def\Dp      {{\ensuremath{\D^+}}\xspace}
\def\Dm      {{\ensuremath{\D^-}}\xspace}

\def\DpDm    {\ensuremath{\Dp {\kern -0.16em \Dm}}\xspace}

\def\B       {{\ensuremath{\PB}}\xspace}

\def\BorBbar {\kern \thebaroffset\optbar{\kern -\thebaroffset \PB}\xspace}

\def\Bd      {{\ensuremath{\B^0}}\xspace}

\def\BdorBdbar {\kern \thebaroffset\optbar{\kern -\thebaroffset \Bd}\xspace}

\def\Bs      {{\ensuremath{\B^0_\squark}}\xspace}

\def\BsorBsbar {\kern \thebaroffset\optbar{\kern -\thebaroffset \Bs}\xspace}

\def\Y#1S{\ensuremath{\PUpsilon{(#1S)}}\xspace}

\def\LorLbar     {\kern \thebaroffset\optbar{\kern -\thebaroffset \PLambda}\xspace}

\def\to                 {\ensuremath{\rightarrow}\xspace}

\newcommand{\as}{{\ensuremath{\alpha_s}}\xspace}

\def\AT#1     {\ensuremath{A_{\mathrm{T}}^{#1}}\xspace}           

\def\C#1      {\ensuremath{\mathcal{C}_{#1}}\xspace}                       
\def\Cp#1     {\ensuremath{\mathcal{C}_{#1}^{'}}\xspace}                    
\def\Ceff#1   {\ensuremath{\mathcal{C}_{#1}^{\mathrm{(eff)}}}\xspace}        
\def\Cpeff#1  {\ensuremath{\mathcal{C}_{#1}^{'\mathrm{(eff)}}}\xspace}       
\def\Ope#1    {\ensuremath{\mathcal{O}_{#1}}\xspace}                       
\def\Opep#1   {\ensuremath{\mathcal{O}_{#1}^{'}}\xspace}                    

\newcommand{\aunit}[1]{\ensuremath{\text{\,#1}}}

\newcommand{\tev}{\aunit{Te\kern -0.1em V}\xspace}
\newcommand{\gev}{\aunit{Ge\kern -0.1em V}\xspace}
\newcommand{\mev}{\aunit{Me\kern -0.1em V}\xspace}
\newcommand{\kev}{\aunit{ke\kern -0.1em V}\xspace}
\newcommand{\ev}{\aunit{e\kern -0.1em V}\xspace}
\newcommand{\gevgev}{\ensuremath{\gev^2}\xspace}
\newcommand{\mevc}{\ensuremath{\aunit{Me\kern -0.1em V\!/}c}\xspace}
\newcommand{\gevc}{\ensuremath{\aunit{Ge\kern -0.1em V\!/}c}\xspace}
\newcommand{\mevcc}{\ensuremath{\aunit{Me\kern -0.1em V\!/}c^2}\xspace}
\newcommand{\gevcc}{\ensuremath{\aunit{Ge\kern -0.1em V\!/}c^2}\xspace}

\def\pb {\aunit{pb}\xspace}
\def\invpb {\ensuremath{\pb^{-1}}\xspace}

\newcommand{\chisq}{\ensuremath{\chi^2}\xspace}
\newcommand{\chisqndf}{\ensuremath{\chi^2/\mathrm{ndf}}\xspace}

\def\deriv {\ensuremath{\mathrm{d}}}

\def\gsim{{~\raise.15em\hbox{$>$}\kern-.85em
          \lower.35em\hbox{$\sim$}~}\xspace}
\def\lsim{{~\raise.15em\hbox{$<$}\kern-.85em
          \lower.35em\hbox{$\sim$}~}\xspace}

\def\pt         {\ensuremath{p_{\mathrm{T}}}\xspace}

\def\ptot       {\ensuremath{p}\xspace}

\def\evtgen     {\mbox{\textsc{EvtGen}}\xspace}

\def\geant      {\mbox{\textsc{Geant4}}\xspace}

\def\herwig     {\mbox{\textsc{Herwig}}\xspace}

\def\photos     {\mbox{\textsc{Photos}}\xspace}
\def\powheg     {\mbox{\textsc{Powheg}}\xspace}
\def\pythia     {\mbox{\textsc{Pythia}}\xspace}

\def\tell1  {TELL1\xspace}
\def\ukl1   {UKL1\xspace}

\newcommand{\lhcborcid}[1]{\href{https://orcid.org/#1}{\hspace*{0.1em}\raisebox{-0.45ex}{\includegraphics[width=1em]{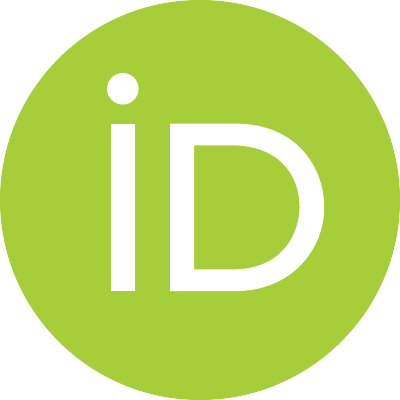}}}}

\hypersetup{
  colorlinks   = true, 
  urlcolor     = blue, 
  linkcolor    = blue, 
  citecolor    = red   
}

\ifEnableSectionTOCLinks
    \usepackage[explicit]{titlesec} 
    
    \let\oldcontentsline\contentsline
    \renewcommand

    \titleformat{\section}{\normalfont\Large\bf}{\hyperlink{tocsection.\thesection}{{\thesection} \parbox[t]{\dimexpr\textwidth-1pc}{#1}}}{1pc}{}

    \titleformat{\subsection}{\normalfont\bf}{\hyperlink{tocsubsection.\thesubsection}{{\thesubsection} \parbox[t]{\dimexpr\textwidth-1pc}{#1}}}{1pc}{}

    \titleformat{name=\section,numberless}[display]{}{}{0pt}{\normalfont\Huge\bfseries #1}
\fi

\usepackage{cite} 
\usepackage{mciteplus}

\usepackage{longtable} 
\usepackage{booktabs}
\usepackage{csquotes}
\usepackage{pgffor}
\usepackage{siunitx}
\usepackage{mathrsfs}
\usepackage{multirow}
\usepackage{listings}

\begin{document}

\newcommand{\mw}{\ensuremath{m_W}\xspace}

\def\DefaultFigureWidth{0.8\linewidth}
\def\MaxFigureWidth{0.99\linewidth}
\def\HalfPageWidth{0.49\linewidth}
\def\Zmumu     {{\ensuremath{\Z \to \mumu}}\xspace}
\def\Ztautau     {{\ensuremath{\Z \to \tautau}}\xspace}
\def\Wtaunu     {{\ensuremath{\W \to \tau\neu}}\xspace}
\def\WW         {{\ensuremath{\Wp\Wm \to \muon X}}\xspace}
\def\ZZ         {{\ensuremath{\Z\Z \to \muon X}}\xspace}
\def\WZ         {{\ensuremath{\W\Z \to \muon X}}\xspace}
\def\Wmn        {{\ensuremath{\W \to \muon\neu}}\xspace}
\def\Wpmn       {{\ensuremath{\Wp \to \mup\neum}}\xspace}
\def\Wmmn       {{\ensuremath{\Wm \to \mun\neumb}}\xspace}
\def\dsigmadpt  {{\ensuremath{\text{d}\sigma/\text{d}\pt}}\xspace}
\def\Deltapt{{\ensuremath{\Delta \pt}}\xspace}
\def\Deltamw{{\ensuremath{\Delta \mw}}\xspace}

\newcommand{\ip}{\ensuremath{\text{IP}}\xspace}
\newcommand{\defiso}{\ensuremath{\mathcal{I}}\xspace}

\newcommand{\runone}{Run~1\xspace}
\newcommand{\runtwo}{Run~2\xspace}
\newcommand{\runthree}{Run~3\xspace}

\newcommand{\bblite}{\ensuremath{\text{BBLite}}\xspace}

\def\dyturbo     {\textsc{DYTurbo}\xspace}
\def\mcfm     {\textsc{MCFM}\xspace}
\def\pythia   {\textsc{Pythia8}\xspace}
\def\photos   {\textsc{Photos}\xspace}
\def\herwig   {\textsc{Herwig}\xspace}
\def\powheg   {\textsc{Powheg}\xspace}
\newcommand{\NumWpCandidates}{27\,586}
\newcommand{\NumWmCandidates}{21\,678}
\newcommand{\NumZCandidates}{3495}
\newcommand{\IntegratedLumiValue}{100}
\newcommand{\IntegratedLumiError}{2}
\newcommand{\FitMinimumDxsWp}{65.8}
\newcommand{\FitMinimumDxsWm}{42.7}
\newcommand{\IntegratedXsecWpValue}{300.9}
\newcommand{\IntegratedXsecWpStatError}{2.4}
\newcommand{\IntegratedXsecWpSystError}{3.8}
\newcommand{\IntegratedXsecWpLumiError}{6.0}
\newcommand{\IntegratedXsecWmValue}{236.9}
\newcommand{\IntegratedXsecWmStatError}{2.1}
\newcommand{\IntegratedXsecWmSystError}{2.7}
\newcommand{\IntegratedXsecWmLumiError}{4.7}
\newcommand{\IntegratedXsecTotalErrorCorrelationWpWm}{0.84}
\newcommand{\mwFitValue}{80366}
\newcommand{\mwFitExperimentError}{130}
\newcommand{\mwFitTheoryErrorBaselinePDF}{37}
\newcommand{\mwFitMinChisq}{25.6}
\newcommand{\gValueWithSystematics}{1.1}
\newcommand{\gErrorWithSystematics}{0.6}
\newcommand{\AvgMwFitValueWithSyst}{80369}
\newcommand{\AvgMwFitTheoryError}{33}
\newcommand{\MomentumSmearingFitChisq}{121.3}
\newcommand{\TotalSigma}{537.8}
\newcommand{\TotalSigmaError}{3.2}
\newcommand{\TotalSigmaLumiError}{7.7}
\newcommand{\mwFitValueNoSyst}{80522}
\newcommand{\mwFitErrorNoSyst}{103}
\newcommand{\mwFitChisquaredNoSyst}{47.0}
\newcommand{\gValueNoSystematics}{1.0}
\newcommand{\gErrorNoSystematics}{0.6}
\newcommand{\FitMisIDRatePionsChiSquareNdof}{8.6/7}
\newcommand{\FitMisIDRateKaonsChiSquareNdof}{8.7/7}
\newcommand{\FitMisIDRateProtonsChiSquareNdof}{13.0/8}
\newcommand{\ZIsoCalibFactor}{0.93}
\newcommand{\ZIsoCalibFactorError}{0.02}
\newcommand{\ZIsoCalibFactorFitMinimum}{16.3}
\newcommand{\QCDIsoCalibFactor}{0.83}
\newcommand{\QCDIsoCalibFactorError}{0.01}
\newcommand{\QCDIsoCalibFactorFitMinimum}{6.12}
\newcommand{\QCDIsoCalibDelta}{0.05}
\newcommand{\QCDIsoCalibDeltaError}{0.01}

\newcommand{\StrongCouplingError}{16}
\newcommand{\qedError}{13}
\newcommand{\qcdError}{14}
\newcommand{\pdfReplicaError}{28}
\newcommand{\avgPdfReplicaError}{22} 
\newcommand{\ChargeIndependentMomentumBiasImpact}{58}
\newcommand{\ChargeDependentMomentumBiasImpact}{41}

\newcommand{\mWCrossCheckImpact}{-3} 
\newcommand{\UnpolarisedXSCrossCheckImpact}{-6} 
\newcommand{\PolarisedXSCrossCheckImpact}{17} 
\newcommand{\FixGCrossCheckImpact}{-58} 
\newcommand{\FloatAlphaCrossCheckImpact}{-1}
\newcommand{\qcdScalesCrossCheckImpact}{-2} 
\newcommand{\qcdCutOffCrossCheckImpact}{3} 
\newcommand{\HigherOrderCrossCheckImpact}{-8} 

\newcommand{\mwPdfShiftUpper}{+14}
\newcommand{\mwPdfShiftLower}{-4}

\newcommand{\mwStatOnlyErrorQuadrature}{106}

\newcommand{\dxsWpMinimum}{131.6}
\newcommand{\dxsWmMinimum}{85.4}


\newcommand{\aMu}{0.119}
\newcommand{\aMuError}{0.007}

\newcommand{\minimumZfit}{10.4/8}
\newcommand{\gZfit}{1.95}
\newcommand{\gZfitError}{0.49}
\newcommand{\asZfit}{0.112}
\newcommand{\asZfitError}{0.006}

\newcommand{\fitmWsuffixsyst}[1]{fixed__floating__fixed__sequential__DYTurbo_03__0__DYTurbo_03_angular_nlo__0__average__#1__FitUnfold__d5TeV__default}

\renewcommand{\thefootnote}{\fnsymbol{footnote}}
\setcounter{footnote}{1}


\begin{titlepage}
\pagenumbering{roman}

\vspace*{-1.5cm}
\centerline{\large EUROPEAN ORGANIZATION FOR NUCLEAR RESEARCH (CERN)}
\vspace*{1.5cm}
\noindent
\begin{tabular*}{\linewidth}{lc@{\extracolsep{\fill}}r@{\extracolsep{0pt}}}
\ifthenelse{\boolean{pdflatex}}
{\vspace*{-1.5cm}\mbox{\!\!\!\includegraphics[width=.14\textwidth]{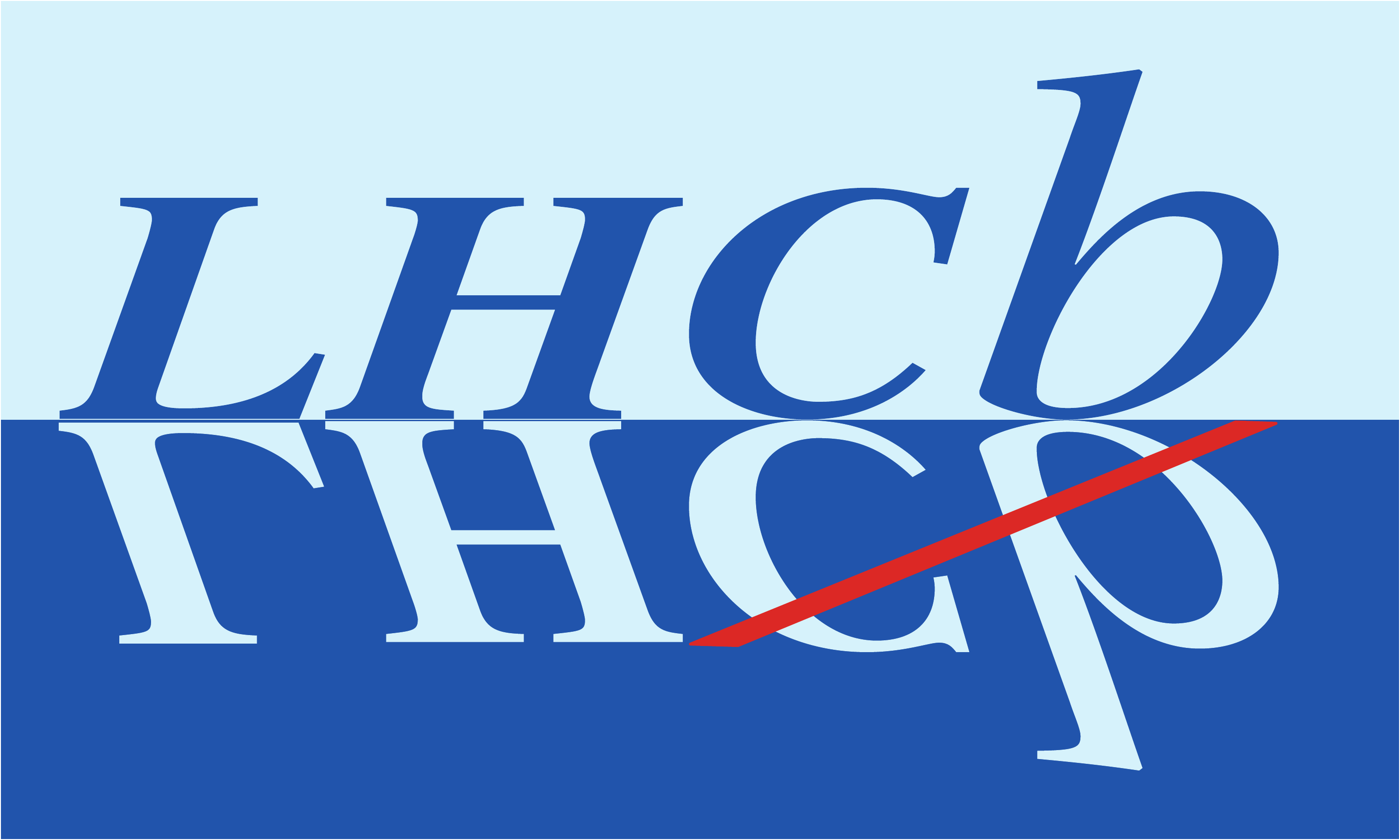}} & &}%
{\vspace*{-1.2cm}\mbox{\!\!\!\includegraphics[width=.12\textwidth]{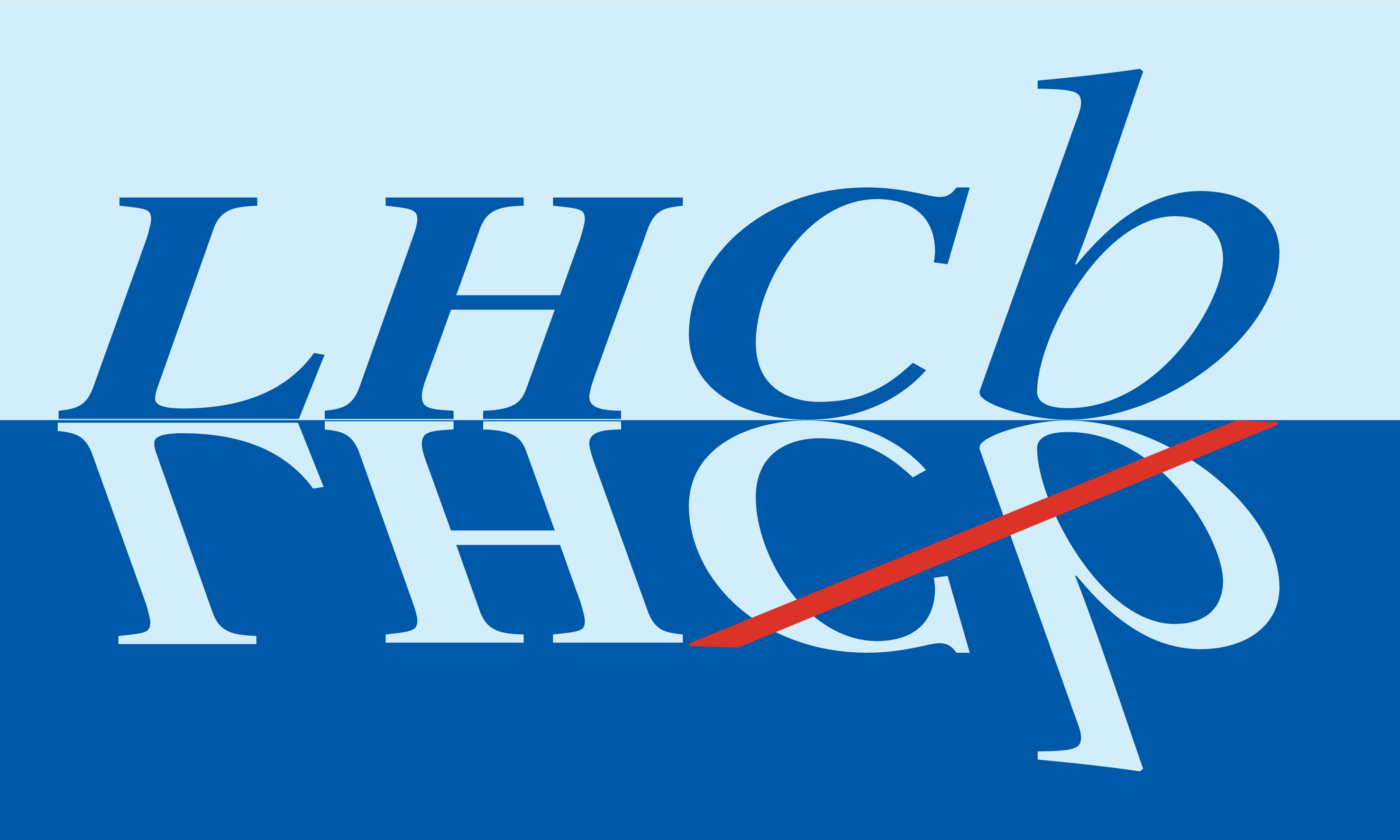}} & &}%
\\
 & & {CERN-EP-2025-197} \\  
 & & {LHCb-PAPER-2025-031} \\  
 & & March 16, 2026 \\ 
 & & \\
\end{tabular*}

\vspace*{1.0cm}

{\normalfont\bfseries\boldmath\huge
\begin{center}
  \papertitle 
\end{center}
}

\vspace*{1cm}

\begin{center}
\paperauthors\footnote{Authors are listed at the end of this paper.}
\end{center}

\vspace{\fill}

\begin{abstract}
\noindent
\noindent The $pp \to W^{\pm} (\to \mu^{\pm} \nu_{\mu}) X$ cross-sections are measured at a proton-proton centre-of-mass energy $\sqrt{s} = 5.02$ TeV using a dataset corresponding to an integrated luminosity of 100 pb$^{-1}$ recorded by the LHCb experiment. Considering muons in the pseudorapidity range $2.2 < \eta < 4.4$, the cross-sections are measured differentially in twelve intervals of muon transverse momentum between $28 < p_\mathrm{T} < 52$ GeV.  
Integrated over $p_\mathrm{T}$, the measured cross-sections are
\begin{align*}
\sigma_{W^+ \to \mu^+ \nu_\mu} &= 300.9 \pm 2.4 \pm 3.8 \pm 6.0~\text{pb}, \\
\sigma_{W^- \to \mu^- \bar{\nu}_\mu} &= 236.9 \pm 2.1 \pm 2.7 \pm 4.7~\text{pb},
\end{align*}
where the first uncertainties are statistical, the second are systematic, and the third are associated with the luminosity calibration. These integrated results are consistent with theoretical predictions.  

This analysis introduces a new method to determine the $W$-boson mass using the measured differential cross-sections corrected for detector effects. The measurement is performed on this statistically limited dataset as a proof of principle and yields
\begin{align*}
m_W = 80369 \pm 130 \pm 33~\text{MeV},
\end{align*}
where the first uncertainty is experimental and the second is theoretical.
\end{abstract}


\begin{center}
  Published in
  JHEP 03 (2026) 148. 
\end{center}

\vspace{\fill}

{\footnotesize 
\centerline{\copyright~\papercopyright. \href{\paperlicenceurl}{\paperlicence}.}}
\vspace*{2mm}

\end{titlepage}


\newpage
\setcounter{page}{2}
\mbox{~}
%
%
%
%


\renewcommand{\thefootnote}{\arabic{footnote}}
\setcounter{footnote}{0}


\cleardoublepage


\pagestyle{plain} 
\setcounter{page}{1}
\pagenumbering{arabic}


\section{Introduction}\label{sec:Introduction}

Massive electroweak vector-boson production is one of the most precisely studied processes in proton-(anti)proton collisions. 
Predictions of the production cross-sections are factorised into hard-process matrix elements and parton distribution functions (PDFs), which describe the momentum distributions of the proton’s quarks and gluons.
At the LHC, the LHCb detector~\cite{LHCb-DP-2008-001,LHCb-DP-2014-002}  fully instruments the forward pseudorapidity region $2 < \eta < 5$. 
Massive vector-bosons with decay products reconstructed in the LHCb acceptance are boosted along the colliding beam axis, corresponding to a large rapidity.
Measurements of the corresponding cross-sections can therefore constrain the PDFs at both high and low parton momentum fractions, complementing the measurements from other collider experiments. 
The LHCb collaboration has measured the \Zmumu cross-section in proton-proton ($pp$) collisions at 
centre-of-mass energies $\sqrt{s}=5.02$, $7$, $8$ and $13\tev$~\cite{LHCb-PAPER-2023-010, LHCb-PAPER-2015-001, LHCb-PAPER-2015-049, LHCb-PAPER-2021-037}, and the \Wmn cross-section at $\sqrt{s}=7$ and $8\tev$~\cite{LHCb-PAPER-2014-033,LHCb-PAPER-2015-049}.
For a given rapidity range, measurements at smaller $\sqrt{s}$ provide constraints on the PDFs at larger momentum fractions than measurements at larger $\sqrt{s}$.
The \Wmn measurements are differential in muon pseudorapidity and integrated over muon transverse momenta $\pt > 20\gev.\footnote{Natural units with $\hbar = c = 1$ are used throughout.}$ 

The production of \PW bosons with decays to electron or muon final states is the basis of the most precise determinations of the \PW-boson mass \mw by the LEP~\cite{LEP3} experiments, the CMS~\cite{CMS:2024}, ATLAS~\cite{ATLAS:mW:2024} and LHCb~\cite{LHCb-PAPER-2021-024} experiments at the LHC, and the CDF~\cite{CDF:2022} and D0~\cite{D0} experiments at the Fermilab Tevatron, with Ref.~\cite{LHC-TeVMWWorkingGroup:2023zkn} reporting a global combination.
The LHC analyses are based primarily on the charged-lepton \pt distribution, while the transverse mass distribution provided the strongest sensitivity in the Tevatron analyses.
All of these measurements are based on fits to the observed number of \PW candidates, in which the model accounts for the detector resolution and efficiency, the background contamination, and the physics of the signal process.
Ideally, the modelling of the physics can be factorised from that of the experiment.
In this case the physics modelling can be reproduced, scrutinised and updated, independently of the experimental collaboration. 

Previous \lhcb\xspace\W-boson cross-section analyses could not be differential in the muon \pt because the shape of the \pt distribution was the basis of the hadronic background subtraction. This paper presents a new measurement of the differential cross-section $\deriv\sigma_\Wmn/\deriv\pt$, following similar measurements by CMS~\cite{CMS:2020cph} and ATLAS~\cite{ATLAS:2025ede} experiments.
For the first time, the cross-section is measured differentially and used in a subsequent determination of \mw within a single analysis, providing a complementary perspective to the traditional direct \pt-fit approach.
The $\dsigmadpt$ measurement is based on a fit, with simulated signal and background templates, to the number of observed signal candidates in intervals of muon \pt and muon isolation, which tends to have larger values for hadronic backgrounds than for the signal. 
The isolation is defined as the scalar sum of the transverse momenta of all other charged particles and electromagnetic calorimeter clusters within an angular separation \mbox{$(\Delta\eta)^2 + (\Delta\phi)^2 < 0.5^2$}, where $\phi$ denotes the azimuthal angle. This definition implies that a well-isolated particle has a small isolation value. The fit simultaneously corrects the data for backgrounds and the detector resolution and efficiency. 
No assumption is made on the shape of the \pt distribution of the signal.

This analysis uses a dataset of $pp$ collisions at \mbox{$\sqrt{s}=5.02\tev$}, recorded during a two-week period in 2017, and corresponding to an integrated luminosity of \mbox{$\mathcal{L}_{\rm int} = \IntegratedLumiValue \pm \IntegratedLumiError \invpb$}~\cite{LHCb-PAPER-2014-047}, acting as a proof-of-principle to measure \mw using this novel approach. The differential cross-section is measured in the fiducial region defined  by $2.2 < \eta < 4.4$, in twelve intervals of transverse momentum spaced equally in the range \mbox{$28 < \pt < 52\gev$}. In this paper, Sec.~\ref{sec:dataset} describes the LHCb detector, the dataset used for this study, and the selection of signal candidates. 
Section~\ref{sec:simulation_corrections} details corrections applied to the simulated samples, while Sec.~\ref{sec:hadronic_background} describes the hadronic background modelling. 
Section~\ref{sec:dxs_measurement} describes the differential cross-section measurement method and results, while Sec.~\ref{sec:integrated_cross_sections} compares the integrated results with predictions.
In Sec.~\ref{sec:mw_fit}, the proof-of-principle \mw determination is presented, while Sec.~\ref{sec:cross-checks} presents a series of cross-checks, before Sec.~\ref{sec:conclusion} concludes this paper.

\section{Detector description and event selection}
\label{sec:dataset}

The \lhcb detector used to collect the data for this analysis is described in detail in Refs.~\cite{LHCb-DP-2008-001,LHCb-DP-2014-002}.
The subdetector systems most relevant for this measurement include a high-precision tracking system consisting of a silicon-strip vertex detector surrounding the $pp$ interaction region~\cite{LHCb-DP-2014-001}, a large-area silicon-strip detector (the TT) located upstream of a dipole magnet with a bending power of about $4{\mathrm{\,T\,m}}$, and three stations of silicon-strip detectors and straw
drift tubes~\cite{LHCb-DP-2017-001} placed downstream of the magnet.
The tracking system provides a measurement of the momentum \ptot of charged particles with a relative uncertainty of around 1\% at $\ptot \approx 200\gev$~\cite{LHCb-DP-2014-002}.
Photons, electrons and hadrons are identified by a calorimeter system consisting of scintillating-pad and preshower detectors, an electromagnetic and a hadronic calorimeter. 
Muons are identified by a system composed of alternating layers of iron and multiwire proportional chambers~\cite{LHCb-DP-2012-002}.
The online event selection is performed by a trigger~\cite{LHCb-DP-2019-001}, which consists of a hardware stage, based on information from the calorimeter
and muon systems, followed by a software stage~\cite{Grieser:2025nch}, which applies a full event reconstruction.

Simulation is required to model and correct for background contributions and the detector response in the \dsigmadpt measurement.
In the simulation, $pp$ collisions are generated using \pythia~\cite{Sjostrand:2007gs} with a specific \lhcb configuration~\cite{LHCb-PROC-2010-056}.
Decays of heavy particles such as \PW bosons, \PZ bosons, and top quarks are modelled with \pythia, which also models final-state photon radiation. 
For lighter particles, their decays are described by \evtgen~\cite{Lange:2001uf}, in which final-state radiation is generated using \photos~\cite{davidson2015photos}.
A hadronic background simulation sample is produced using the hard-QCD processes in \pythia with a minimum transverse momentum of $18\gev$.
The interaction of the generated particles with the detector and its response are implemented using the \geant toolkit~\cite{Allison:2006ve, *Agostinelli:2002hh} as described in Ref.~\cite{LHCb-PROC-2011-006}. 

This analysis uses events selected by the hardware trigger requiring a muon with $\pt > 5\gev$.
No attempt is made to reconstruct the neutrino from an imbalance of transverse momentum in the detector, while all identified muons with $\pt > 28\gev$ and $2.2 < \eta < 4.4$ are considered as \Wmn signal candidates.
The $\eta$ region is slightly restricted compared to the detector coverage, as in Ref.~\cite{LHCb-PAPER-2021-024}, to ensure that the isolation is well measured.
The background from \Zmumu decays is suppressed by roughly a factor of two, with a negligible inefficiency for the signal, by rejecting events in which there is a second muon with $\pt > 25\gev$ and $2.0 < \eta < 4.5$.
Candidates must have a reasonable probability of originating from a primary $pp$ interaction vertex, as reconstructed by the vertex detector,
which suppresses the heavy-flavour hadron background.
A requirement on the track-fit \chisqndf suppresses the background from muonic decays of light long-lived hadrons.
To qualify as signal candidates, the isolation value must be under $8\gev$. This criterion helps mitigate the background from hadrons, whose isolation values are around $\mathcal{O}(10\gev)$, whereas muons originating from \PW- and \PZ-boson decays generally have $\mathcal{O}(1\gev)$ isolation values.

Candidate \Zmumu decays, reconstructed from oppositely charged muons, are used in several calibrations. They must have a dimuon mass between $77$ and $105\gev$, and the muons must satisfy $\pt > 20\gev$ and $1.7 < \eta < 4.4$, in addition to the same track-fit $\chisqndf$ requirement imposed on the signal candidates.
The minimum allowed $\eta$ value is smaller than the value of $2.2$ used in the selection of \PW-boson signal candidates.
This choice increases the number of muons from \PZ-boson decays with $\eta > 2.2$ that are used for calibrations.
Looser requirements are also imposed on the isolation and the compatibility of the tracks with a primary $pp$ interaction vertex.

\begin{figure}
    \centering
    \includegraphics[width=\DefaultFigureWidth]{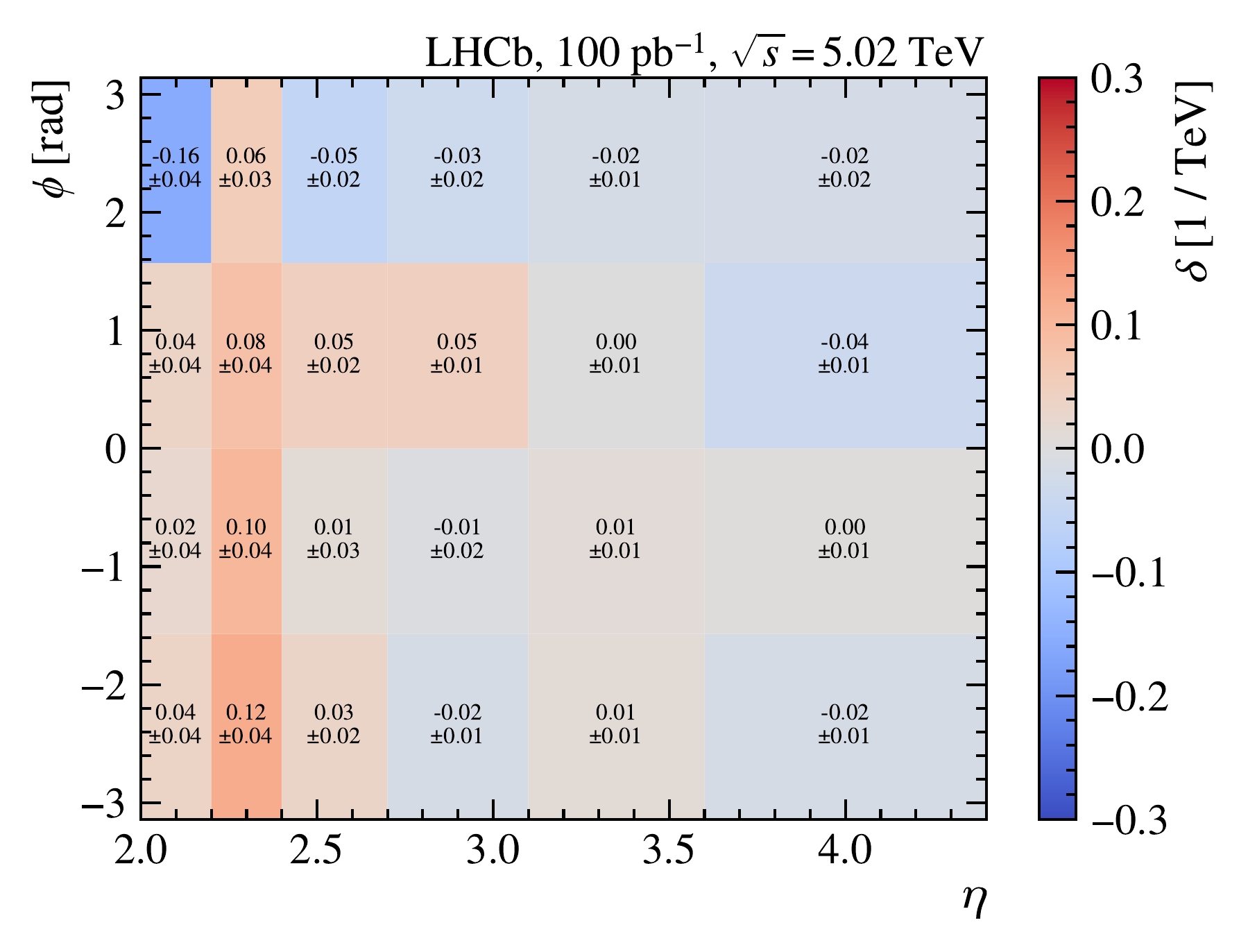}
    \caption{Biases in $q/p$ estimated with the pseudomass method in $(\eta, \phi)$ intervals.}
    \label{fig:pseudomass_corrections}
\end{figure}

Muons from \PW- and \PZ-boson decays tend to have higher momenta than charged particles from heavy-flavour hadron decays. Thus, they are more susceptible to charge ($q$) dependent curvature biases, of the form $q/p \rightarrow q/p + \delta$, caused by residual misalignment effects. 
Furthermore, the $5.02\tev$ data sample used in this analysis was recorded for only one polarity of the dipole magnet, which makes this analysis susceptible to residual curvature biases.
The pseudomass method~\cite{Barter:2021npd,LHCb-DP-2023-001} is used to determine the curvature bias corrections to be applied to the muons in the data. 
In this method, estimates of the dimuon mass in $\PZ \to \mu^+\mu^-$ decays are made
using only the momentum of muons of a given charge, allowing charge-dependent biases to be probed.
The asymmetry in the peak positions of the distributions for the two charges can be translated into estimates of the $\delta$ biases.
In Ref.~\cite{LHCb-DP-2023-001}, the application of this method to the full LHCb Run-2 dataset is described, where 
biases of up to $\mathcal{O}(0.1\tev^{-1})$ are seen in the data while the simulation only exhibits very small asymmetries associated to the mixture of vector and axial-vector couplings in the \Zmumu process.
Applying this method to the $5.02\tev$ sample, Fig.~\ref{fig:pseudomass_corrections} shows, in intervals of muon $\eta$ and $\phi$, the biases in data, having also subtracted the small biases observed in the simulation.
Applying these as curvature corrections to the muons in data results
in an improvement of $\mathcal{O}(10\%)$ in the \Zmumu mass resolution and greatly reduces the required size and complexity of the simulation corrections described in the following section.

Following the curvature bias corrections and the selections described above, 
$\NumWpCandidates$ \mbox{\Wpmn} candidates and $\NumWmCandidates$ \mbox{\Wmmn} candidates are retained for the \dsigmadpt measurement, and $\NumZCandidates$ \Zmumu candidates are retained for data-driven adjustments to the simulation that are described in the following section.

\section{Corrections to the simulation}
\label{sec:simulation_corrections}

The simulation is used to model the muon detection efficiency and backgrounds, and subsequently to correct the data for these contributions.
Corrections to the simulation are required to improve the accuracy of this model in describing the data.

\subsection{Muon momentum smearing} 
\label{sec:smearing}

Although the pseudomass corrections improve the momentum resolution in data, some effects contributing to the momentum resolution are still underestimated in the simulation. Therefore, a smearing of the muon momenta in the simulation is required.
In the simulation, the muon momenta are multiplied by a factor 
\begin{equation}\label{eq:smearing}
  (1 + \alpha)(1 + f_1(\eta)\mathcal{R}_1\sigma_1)(1 + pf_2(\eta)\mathcal{R}_2\sigma_2), 
  \end{equation}
where $\mathcal{R}_{1,2}$ are independent random numbers sampled from a standard normal distribution, $\alpha$ is a momentum scale offset, and $\sigma_1$ and $\sigma_2$ are momentum and curvature smearing parameters, respectively. 
The function $f_1(\eta)$ is defined to be 1.0 for $\eta < 3.3$ and 1.5 for $\eta \geq 3.3$, while $f_2(\eta) = 1/\cosh\eta$.

\begin{figure}[tbp]
  \centering
  \includegraphics[width=\DefaultFigureWidth]{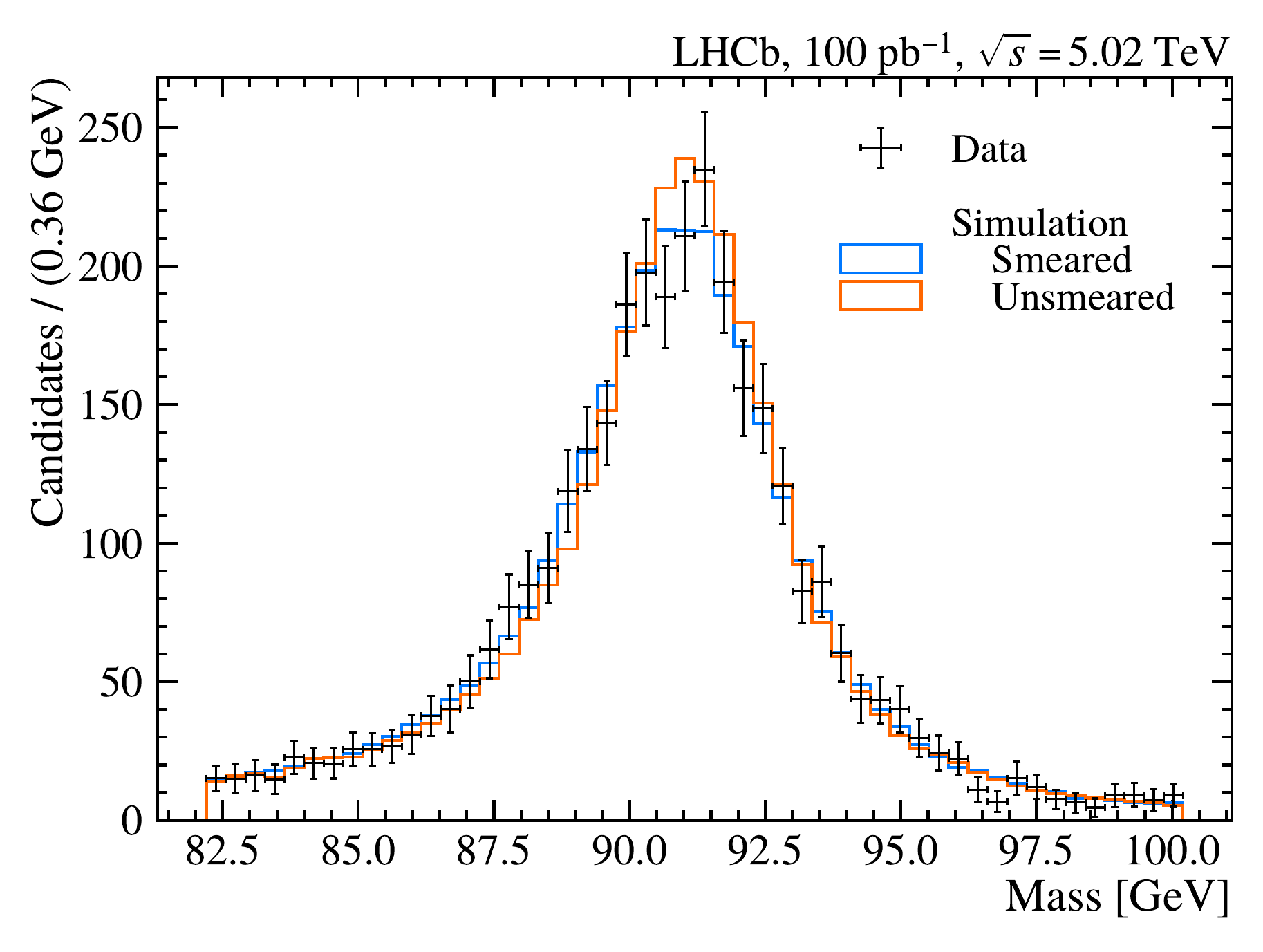}
  \caption{Mass distribution of the $Z \to \mu^+ \mu^-$ candidates. The simulation before and after the application of the momentum smearing is also shown.}
  \label{fig:MomentumSmearingFit}
\end{figure}

The values for the $\alpha$ and $\sigma_2$ parameters are obtained by minimising the $\chisq$ between the \Zmumu mass distribution in data and that in the simulation, where the simulation is smeared according to these parameters.
Independent parameters are used for muons in the regions $\eta < 2.2$ and $2.2 < \eta < 4.4$.
The \Zmumu candidates are categorised as having zero, one or two muons in the $2.2 < \eta < 4.4$ region, which yields three separate dimuon mass distributions.
The background is approximately three orders of magnitude smaller than the signal and can be safely neglected in the fit. 
The value of $\sigma_1$, which has minimal influence in this analysis, is fixed at $2\times 10^{-3}$, similar to what was chosen for the previous \mw analysis at LHCb~\cite{LHCb-PAPER-2021-024}.
The $\chisq$ of the fit is $\MomentumSmearingFitChisq$ for 146 degrees of freedom, and the best-fit values of the other parameters are presented in Table~\ref{tab:MomentumSmearingFit}. 
Figure~\ref{fig:MomentumSmearingFit} shows the mass distribution in data, compared to the simulation before and after the smearing with the best-fit values of the smearing parameters.
While the momentum smearing only has a modest effect on the modelling of the \Zmumu mass distribution in this small dataset, it is important to estimate and propagate the statistical uncertainties on the smearing parameters.

\begin{table}[tb]\centering
\caption{\label{tab:MomentumSmearingFit} Results of the momentum smearing fit.}
\begin{tabular}{lll}
Parameter & $\eta$ range & Fit result \\
\hline
\multirow{2}{*}{$\alpha$}   
  & $\eta < 2.2$       & $(\phantom{-}0.2\pm 0.9) \times 10^{-3}$ \\
  & $2.2 < \eta < 4.4$ & $(-1.2\pm 0.6) \times 10^{-3}$ \\
\multirow{2}{*}{$\sigma_2$} 
  & $\eta < 2.2$       & $(~15.4\pm 3.1) \times 10^{-5}$ GeV$^{-1}$ \\
  & $2.2 < \eta < 4.4$ & $(~24.7\pm 2.8) \times 10^{-5}$ GeV$^{-1}$ \\
\end{tabular}

\end{table}

\subsection{Muon detection efficiency}

Following the approach of Ref.~\cite{LHCb-PAPER-2021-024}, the muon trigger, tracking, and identification efficiencies are measured from the \Zmumu samples in both the data and simulation. 
One muon, denoted as the ``tag", must satisfy the requirements of all stages of the trigger system. 
The fraction of candidates in which the other muon, the ``probe", also satisfies the trigger requirements provides an estimate of the per-muon trigger efficiency. 
Similarly, a dedicated sample, without a muon-identification requirement applied to the probe, is used to estimate the corresponding efficiency. 
The track-reconstruction efficiency is estimated with a sample in which the probe muon is reconstructed only using hits in the muon system and the TT chambers.\footnote{The standard charged-particle reconstruction does not explicitly require TT clusters.}

The muon trigger, tracking and identification efficiencies are estimated in five intervals of $\eta$ in both data and simulation.
Weights are then evaluated as the ratios of data to simulation, and used to correct for the differences in the simulated efficiencies.
Identification and tracking efficiency corrections are typically within $\mathcal{O}(1\%)$ of unity for all $\eta$ intervals.
The trigger efficiency corrections are similarly small at low $\eta$ but in the highest $\eta$ interval, with $3.85 < \eta < 4.40$, the correction factor is around $0.85$.

\subsection{Muon-isolation calibration}

\begin{figure}[tbp]
    \centering
    \includegraphics[width=\DefaultFigureWidth]{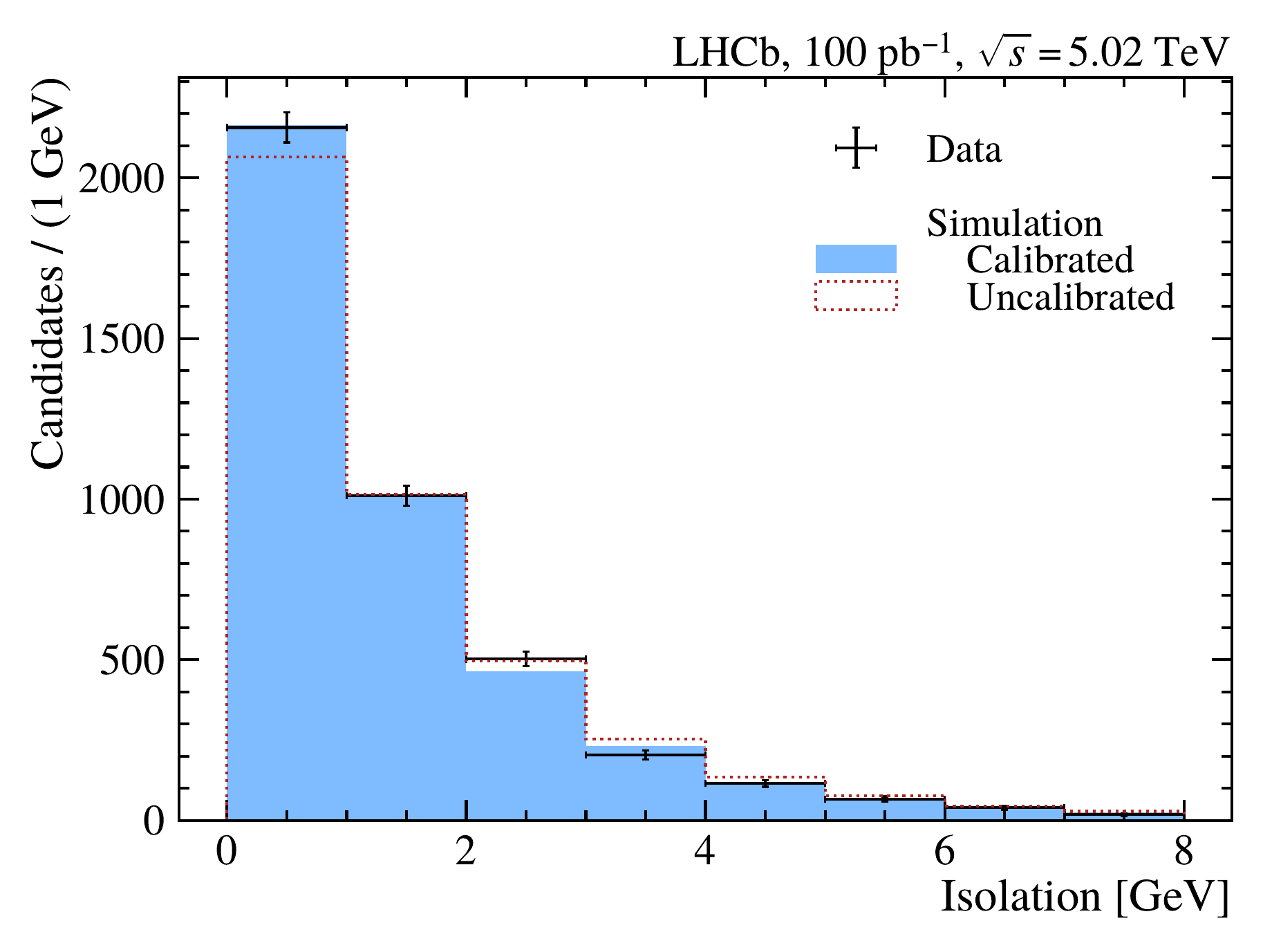}
    \caption{Muon-isolation distribution of the $Z \to \mu^+ \mu^-$ candidates. The simulation before and after the application of the isolation calibration is also shown.}
    \label{fig:iso_calibration_signal}
\end{figure}

Figure~\ref{fig:iso_calibration_signal} shows the isolation variable distribution for the  \Zmumu sample, including muons of both charges, compared to the simulation. 
The modelling of this distribution depends on both the detector response and the physics processes, including the hadronic recoil in the hard process and the underlying event. 
To account for inaccuracies in the simulation, a multiplicative calibration factor of $\ZIsoCalibFactor \pm \ZIsoCalibFactorError$ is determined by fitting the simulated isolation distribution to that observed in \Zmumu data, as shown in Fig.~\ref{fig:iso_calibration_signal}. The fit yields a minimum \chisqndf of $\ZIsoCalibFactorFitMinimum/13$. This calibration factor is then applied to the isolation values of muons in the \Wmn signal simulation.

\section{Hadronic-background modelling}
\label{sec:hadronic_background} 

One of the largest sources of background among the \Wmn candidates consists of light charged hadrons, which are misidentified as muons due to decays in flight or particles in hadronic showers that are not fully absorbed before the muon system.
The majority of the hadronic background is where the muon candidate is actually a charged pion or kaon, with a smaller contribution coming from protons.

\begin{figure}
    \centering
    \includegraphics[width=\DefaultFigureWidth]{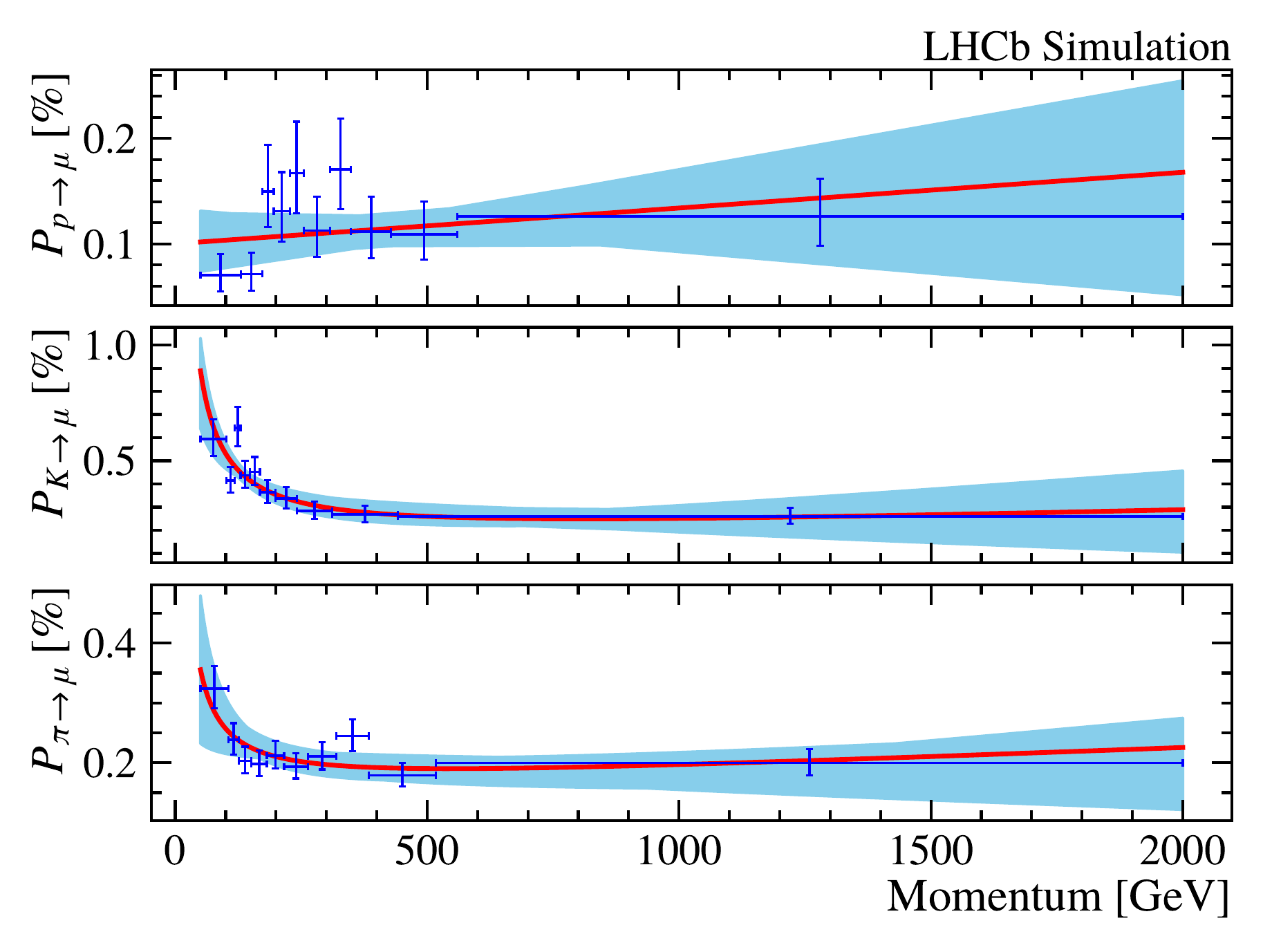}
    \caption{Momentum-dependent probabilities for (lower) pion, (central) kaon and (upper) proton misidentification as muons in the simulation. The fit results are also shown and are represented by the red lines. The blue bands show the area covered by the systematic uncertainties. Note that the misidentification probabilities correspond to the muon-identification requirements used in this analysis, which are not optimised for low momentum.}
    \label{fig:hadron-mis-id}
\end{figure}

Figure~\ref{fig:hadron-mis-id} shows the probabilities for pions, kaons and protons to be misidentified as muons in the hadronic-background simulation sample. 
For the pions and kaons, the momentum dependence is parametrised as the sum of two components.
The first accounts for decays to muons and has the form $1 - \exp\left(-ml / p\tau\right)$ for a particle of mass $m$ and lifetime $\tau$, where $l=\SI{15}{\metre}$ is the approximate length of the detector up to the second muon station.
The second, which increases linearly with momentum, accounts for hadronic shower particles penetrating through the calorimeters or for decays of nearby hadronic particles to muons.
For protons, only the second term is included.
The results of the fits of these functions to the simulated probabilities are shown in Fig.~\ref{fig:hadron-mis-id}. The \chisqndf fit values for pions, kaons, and protons are $\FitMisIDRatePionsChiSquareNdof$, $\FitMisIDRateKaonsChiSquareNdof$, and $\FitMisIDRateProtonsChiSquareNdof$, respectively.
For pions and kaons, the probabilities vary between $\mathcal{O}(10^{-3})$ and $\mathcal{O}(10^{-2})$, while for protons they are $\mathcal{O}(10^{-3})$.
The parametrisations of the misidentification rates are used to assign weights that replace explicit muon-identification requirements in the simulation of the hadronic background.

\begin{figure}[tbp]
    \centering
    \includegraphics[width=\DefaultFigureWidth]{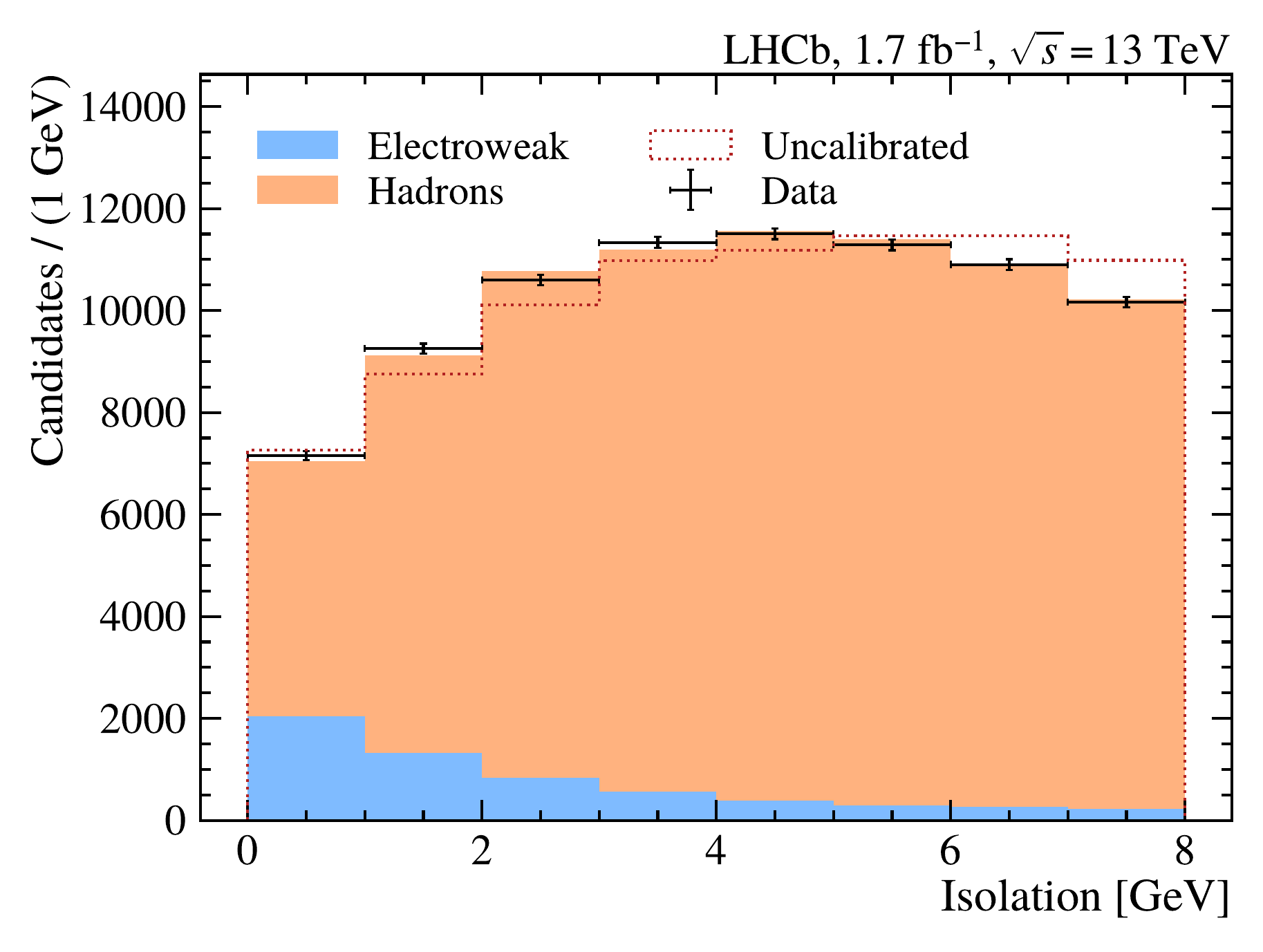}
    \hfill
    \caption{Muon-isolation distribution in the hadron-enriched sample, with both charges combined. The simulation before and after the application of the isolation calibration are also shown.}
    \label{fig:iso_calibration_hadrons}
\end{figure}

To study high-\pt hadron behaviour in the data, a sample enriched in high-\pt hadrons is required. A sample is selected from the data recorded in $pp$ collisions at $\sqrt{s}=13\tev$ in 2017.
Since no dedicated hadron trigger was available during the $\sqrt{s}=5.02\tev$ run, the $13\tev$ sample is used to calibrate the modelling of the hadronic background at $5.02\tev$.
In the software trigger, the hadron candidates are selected in the same way as the signal candidates, except that the muon-identification requirement is replaced by a randomised event selection to reduce the rate.
Figure~\ref{fig:iso_calibration_hadrons} shows the isolation distribution of these candidates, with the same kinematic and track-quality requirements applied as for the \Wmn selection. 
Similarly to the muons from electroweak processes, a calibration is applied to the simulation to account for possible inaccurate modelling.
A simultaneous fit is performed on the isolation distributions of the positively and negatively charged hadrons. 
The data are described by the sum of a small electroweak component at lower isolation values and the main hadron component, with simulation.
The electroweak component is predominantly hadronic decays of $\tau$ leptons from \PW- and \PZ-boson decays. 
In the fit, the hadronic fraction for each charge is allowed to vary freely. The isolation values for the hadrons of charge $q$ are scaled by a calibration factor of $f + q\zeta$, with $f$ and $\zeta$ varying freely in the fit. 
The fit has a minimum $\chisqndf$ of $8.2/10$ and the best-fit calibration parameters are $f=\QCDIsoCalibFactor \pm \QCDIsoCalibFactorError$ and $\zeta= \QCDIsoCalibDelta \pm \QCDIsoCalibDeltaError$.
Figure~\ref{fig:iso_calibration_hadrons} shows the distribution in data and the fit results.

\section{Differential cross-section measurement}
\label{sec:dxs_measurement}

\subsection{Likelihood function}

The differential cross-sections \dsigmadpt are determined from a fit to the two-dimensional distributions of the \Wmn candidates, corresponding to the product of twelve \pt intervals in the range \mbox{$28 < \pt < 52\gev$} and eight intervals for the isolation below $8\gev$. Each \pt interval has a width of $2\gev$, which is denoted $\Deltapt$.
In data, the numbers of candidates in the 96 intervals $i$ of \pt and isolation are denoted $k_i$.
The fit to these data minimises the function
\begin{equation}
    -\ln{\mathscr{L}} = \sum_i^n \left[-k_i \ln{\beta_i \lambda_i} + \beta_i \lambda_i \right] + \frac{1}{2}\sum_{i,k}^n (\beta_i - 1)\left(\frac{C_{i,k}}{\lambda_i\lambda_k}\right)^{-1}(\beta_k - 1),
    \label{eq:BBL_likelihood}
\end{equation}
where $\lambda_i$ are the model predictions based on simulation; $\beta_i$ are nuisance parameters accounting for the finite sizes of the simulation samples; and the covariance matrix $C$ represents the total statistical uncertainty in the predictions, stemming from the limited sizes of the simulation samples. The first term in Eq.~\ref{eq:BBL_likelihood} represents the contribution from the Poisson probability of observing the data given the expected counts. The $\beta$ values are minimised via an analytic solution~\cite{BeestonBarlow:1993,Conway:2011in}.
The model predictions are composed as 
\begin{equation}
    \lambda_i = \lambda_i^{\rm s} + \lambda_i^{\rm had} + \sum_b \lambda_i^{b},
\end{equation}
where the first term corresponds to the signal, the second to the hadronic background, and the third to all other backgrounds. 

\subsection{Signal component}

The signal component is predicted as
\begin{equation}
\lambda^{\rm s}_i(\dsigmadpt) = \mathcal{L}_{\rm int} \sum_j R_{ij} \dsigmadpt_j \Deltapt,
\end{equation}
where $R_{ij}$ is a response matrix, defined as 
\begin{equation}
R_{ij} = \frac{n^{\rm rec}_{ij} \bar{w}_{ij}}{n^{\rm gen}_j},
\end{equation}
where $n^{\rm rec}_{ij}$ is the number of simulated decays with a true \pt corresponding to the interval $j$ and being reconstructed in interval $i$, while $\bar{w}_{ij}$ is the corresponding average efficiency weight. 
The denominator ${n^{\rm gen}_j}$, which is the number of events generated with the true \pt in interval $j$, explicitly eliminates the dependence of this analysis on any assumption about \dsigmadpt.\footnote{Up to any variation in the detection efficiency within the $2\gev$ intervals of \pt.}

The response matrix has 14 intervals along the true \pt axis, where twelve of these exactly match the reconstructed \pt intervals. 
The other two, referred to as the ``underflow" and ``overflow", respectively, correspond to  \mbox{$26 < \pt < 28\gev$} and \mbox{$52 < \pt < 54\gev$}.
These are effectively treated as two extra background components in the fit, but with a special treatment of their normalisations. Both cross-sections are predicted as the sum of the twelve cross-sections spanning the fiducial region, which vary freely in the fit, multiplied by factors determined from the simulation, which are fixed in the fit. With this strategy there are no freely varying parameters associated to the underflow and overflow cross-sections, and no assumption is required for their {\em absolute} values; only two {\em relative} cross-sections.

\begin{figure}
    \centering
    \includegraphics[width=\DefaultFigureWidth]{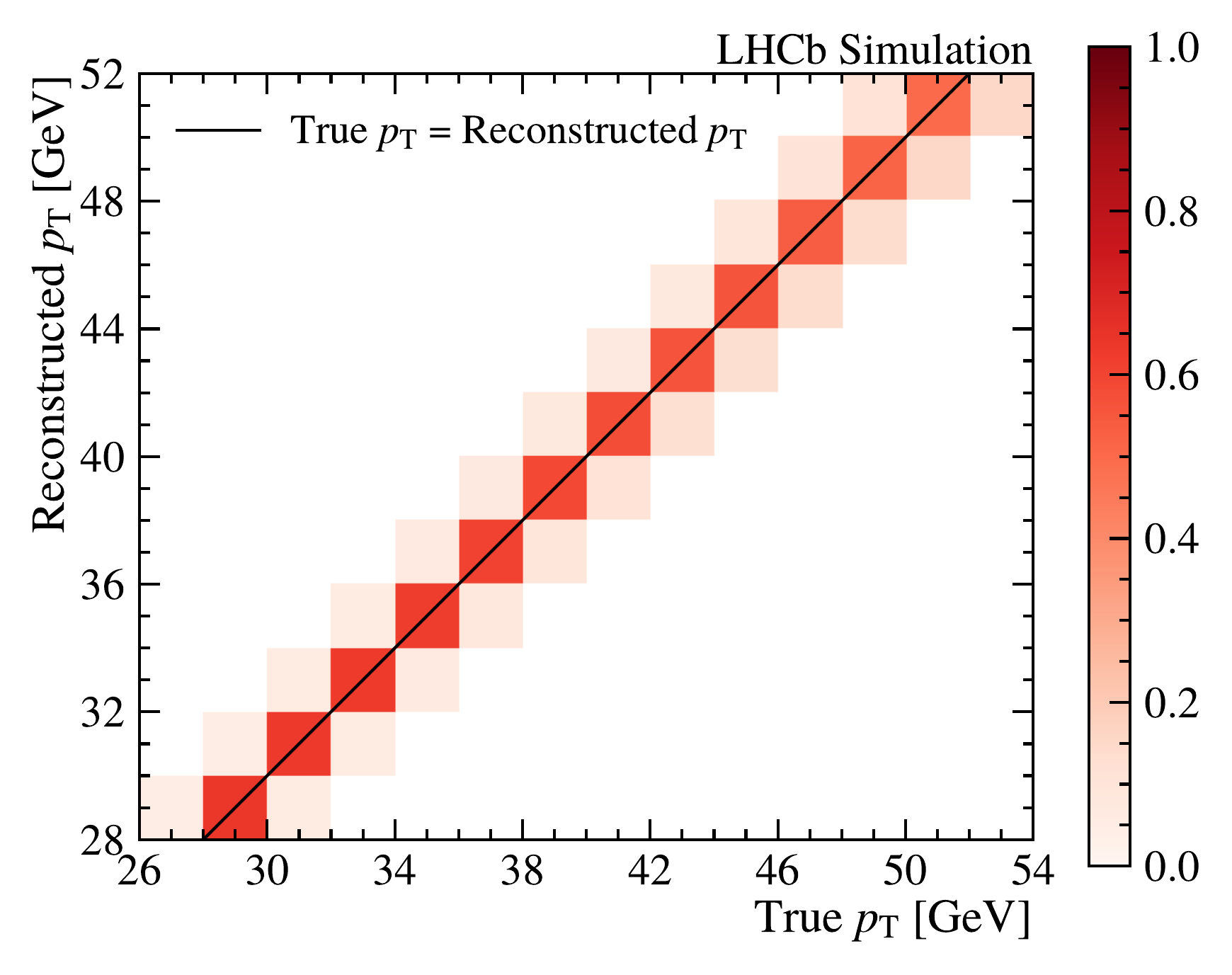}
    \caption{Response matrix for the $W^+$ decay, after integration over the isolation intervals.}
    \label{fig:response_matrix_Wp}
\end{figure}

Figure~\ref{fig:response_matrix_Wp} shows the response matrix for the \Wp decay process, having integrated over the isolation. 
A very similar response matrix is obtained for the \Wm case.
Roughly 70\% of the signal candidates have the true and reconstructed \pt in the same interval, which is expected comparing the $2\gev$ interval width with the resolution of $\mathcal{O}(1\gev)$.
Statistical uncertainties and correlations from the response matrix are appropriately propagated, taking into account its multinomial nature.

\subsection{Backgrounds}

The light hadron background is modelled using a normalised template $t_{{\rm had}, i}^{\rm rec}$, derived from simulation after the misidentification probability weights are applied to reconstructed hadrons. This template represents the expected shape of the background across bins, with $\sum_i t_{{\rm had}, i}^{\rm rec} = 1$. The expected number of background events in bin $i$ is expressed relative to the observed number of signal candidates $n^{\rm obs}$ as
\begin{equation}
\lambda_i^{\rm had} = n^{\rm obs} f_{\rm had} t_{{\rm had}, i}^{\rm rec},
\label{eq:qcd_bkg_original_model}
\end{equation}
where $f_{\rm had}$ is a freely varying fraction.

The remaining backgrounds from \Zmumu, \Ztautau, $\PW \to \tau\nu$ and decays of heavy-flavour hadrons and top quarks are modelled using simulated samples. For each background $b$, $n^{\rm gen}_b$ events are generated, resulting in a simulated distribution $n^{\rm rec}_{b,i}$.
The associated fit model is 
\begin{equation}
\lambda^b_i = \left(\frac{n^{\rm obs}_Z}{\sigma_Z \varepsilon_Z}\right) \left( \frac{n^{\rm rec}_{b,i}}{n^{\rm gen}_b}\right) \sigma_{b} ,
\end{equation}
where $\sigma_b$ is the background process cross-section. 
The first factor is an estimate of the integrated luminosity based on the observed number of \Zmumu events $n^{\rm obs}_Z$, the corresponding cross-section $\sigma_Z$ and the simulated efficiency $\varepsilon_Z$. 
The cross-sections are evaluated at $\mathcal{O}(\alpha_s^2)$ using the \mcfm~\cite{MCFM} program with the NNPDF4.0 parton distribution functions.\footnote{In this paper $\alpha_s$ denotes the strong coupling evaluated at a scale corresponding to the \PZ-boson mass.}
The backgrounds from heavy-flavour hadron decays are included with the leading-order cross-section computed using \pythia.

\subsection{Fit results}

\begin{figure}[tbp]\centering        
\includegraphics[width=\DefaultFigureWidth]{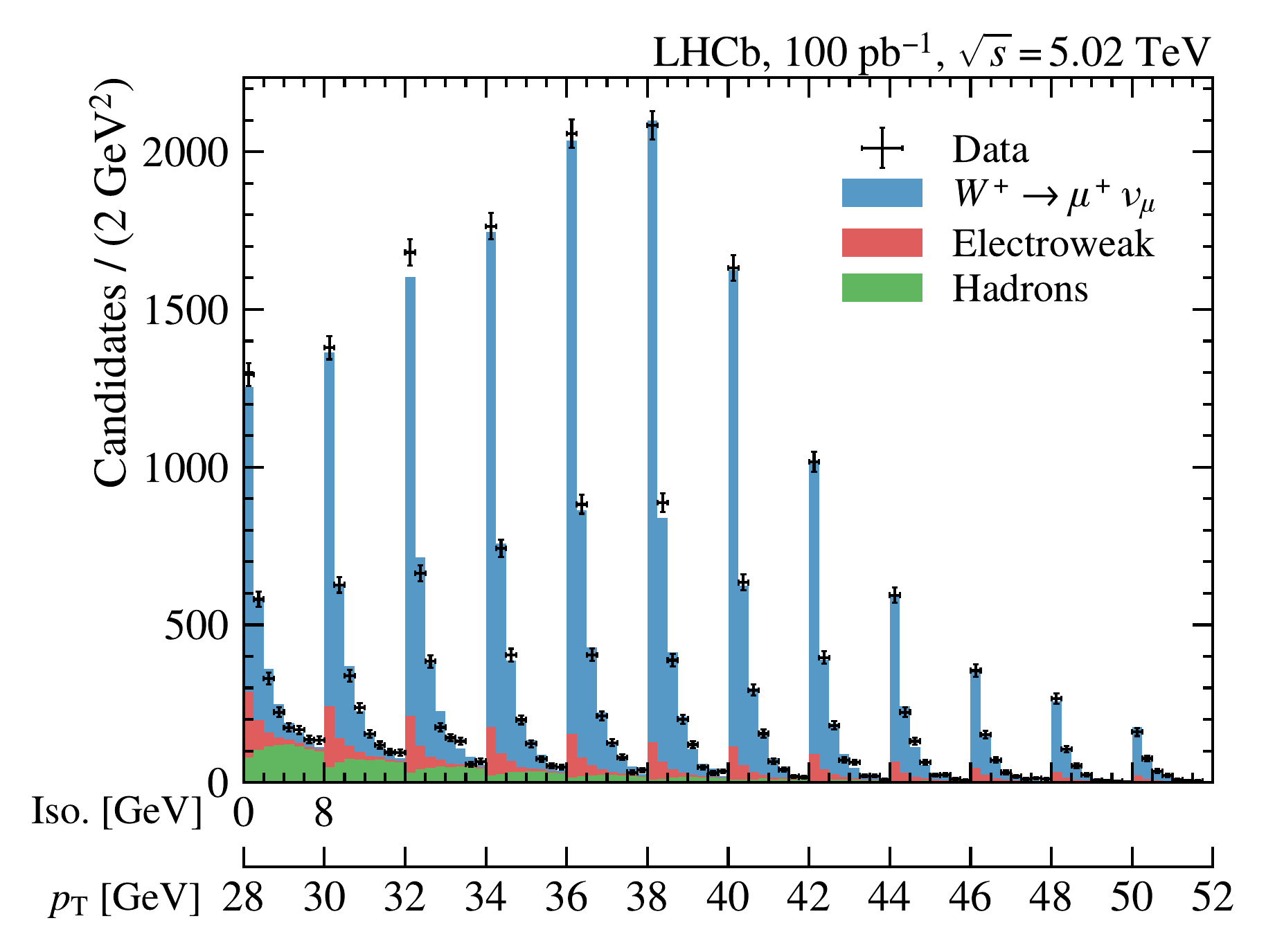}
\includegraphics[width=\DefaultFigureWidth]{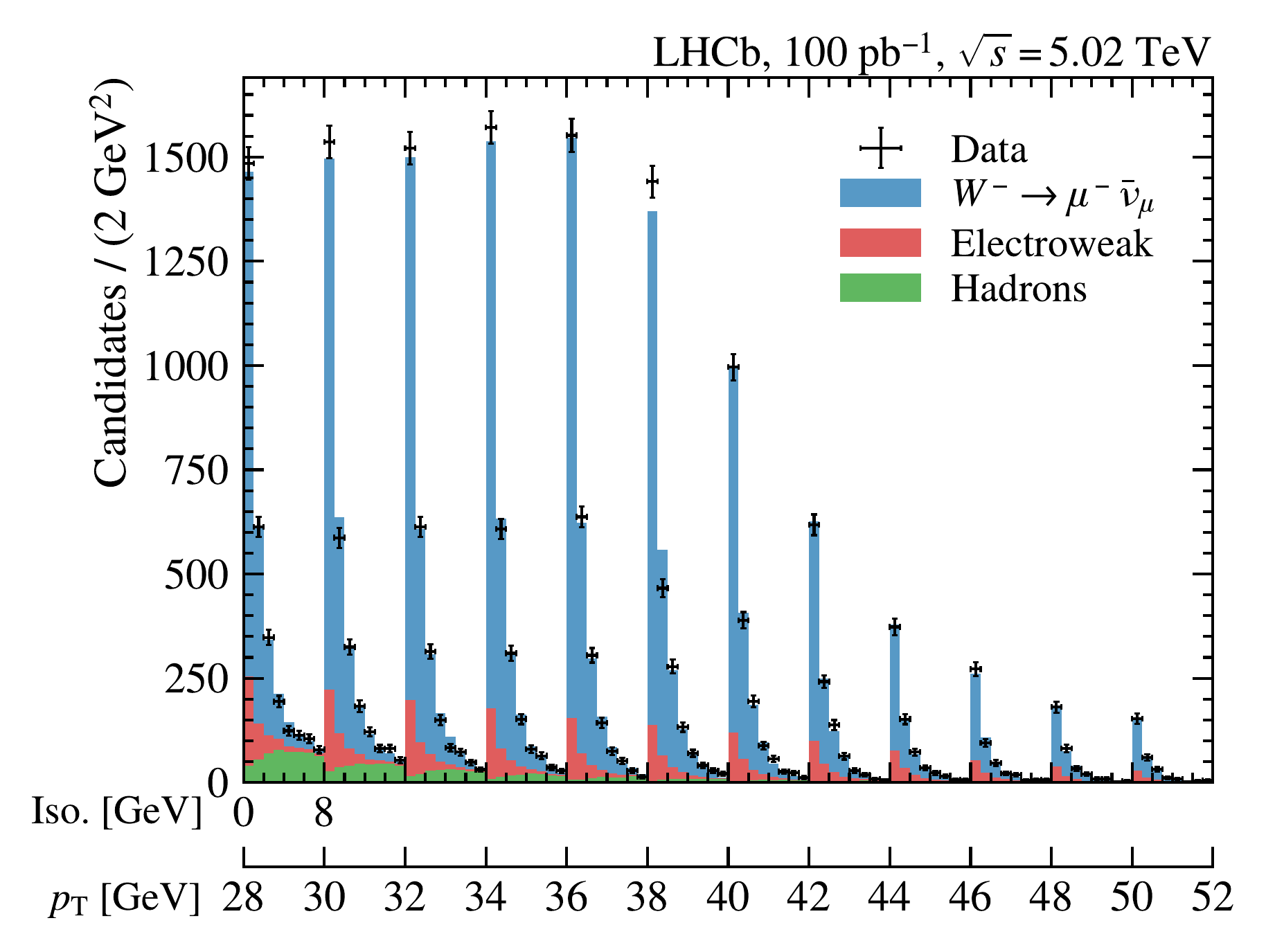}
\caption{Distributions of $p_\mathrm{T}$ and $\mathrm{isolation}$ for the (upper) $W^+$ and (lower) $W^-$ candidates. Eight intervals of isolation up to $8\,\text{\,Ge\kern-0.1em V}
$ are repeated over the twelve $p_\mathrm{T}$ intervals. The results of the differential cross-section fits are also shown, where the electroweak background component includes the relatively small contribution from heavy-flavour hadron decays.}
\label{fig:fit_results}
\end{figure}

\begin{figure}\centering
    \includegraphics[width=\DefaultFigureWidth]{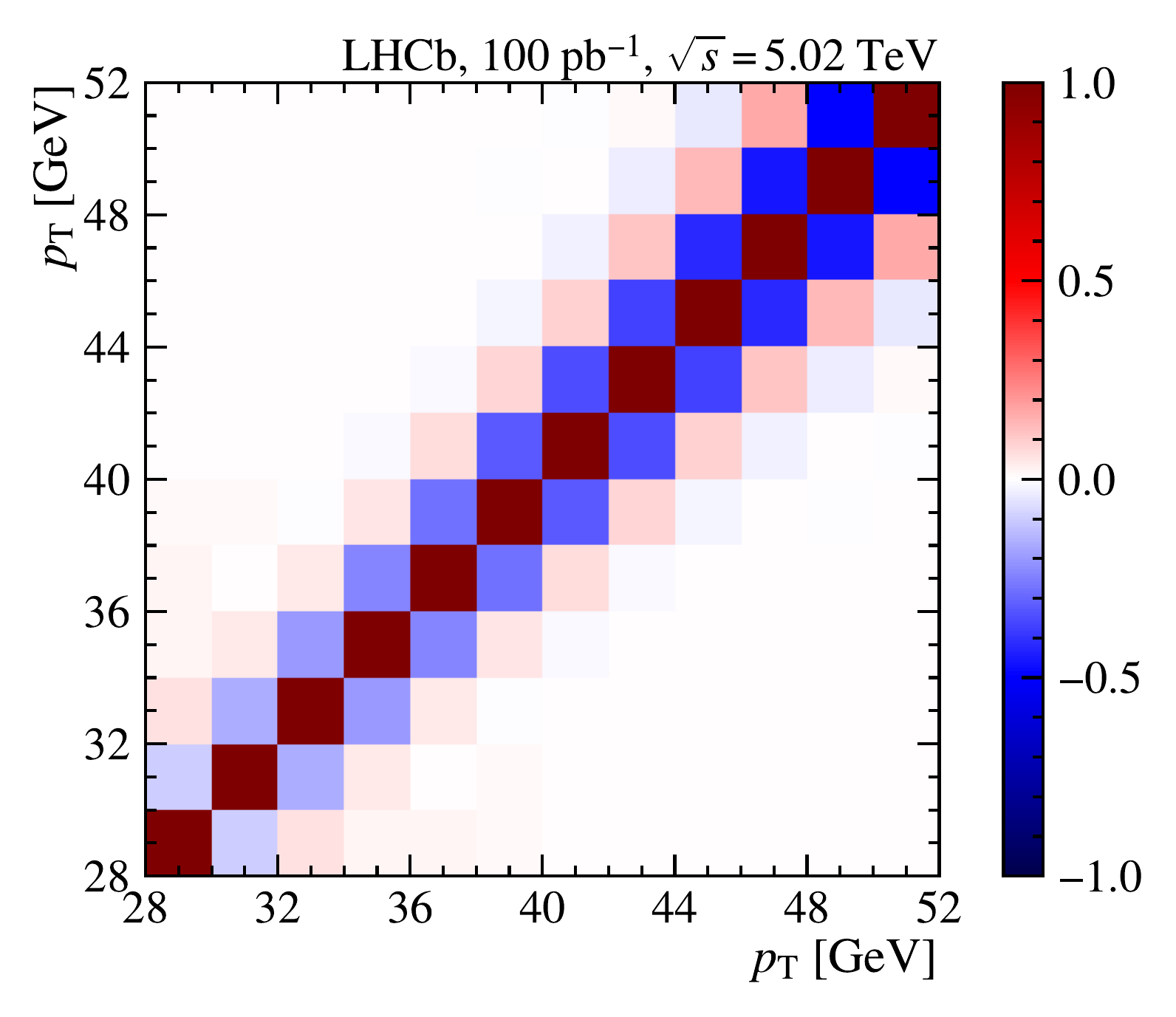}
    \caption{Correlation matrix corresponding to the statistical uncertainties on the ${\ensuremath{\text{d}\sigma/\text{d}p_\mathrm{T}}}$ fit results for the $W^+$ process. The same qualitative patterns are seen for the equivalent figure for the other $W$-boson charge.}
    \label{fig:fit_correlation_matrix}
\end{figure}

Figure~\ref{fig:fit_results} shows the $(\pt, \mathrm{isolation})$ distributions and the fit results, which are independent for the \Wp and \Wm distributions. 
In each case, there are $12 \times 8$ intervals, and there are $12$ freely varying \dsigmadpt values and 1 freely varying hadronic background parameter, corresponding to 83 degrees of freedom.
The fit minima correspond to \chisq values of $\dxsWpMinimum$ and $\dxsWmMinimum$ for the \Wpmn and \Wmmn fits, respectively.\footnote{The \chisq calculation accounts for the constant terms in the log-likelihood that are commonly omitted since they do not depend on the fit parameters.}
The best-fit \dsigmadpt values and the statistical uncertainties are presented in Table~\ref{tab:combined_xs_results}.
Figure~\ref{fig:fit_correlation_matrix} shows the correlation matrix corresponding to the statistical uncertainties for the fit to the \Wpmn data.
Moving away from the diagonal, a pattern of alternating negative and positive correlations is seen but this is damped to the $\mathcal{O}(1\%)$ level after two cells away from the diagonal.
The corresponding matrix for the other \PW-boson charge is similar.

\begin{table}
    \centering
    \caption{Results of the differential cross-section fits, where the first and second uncertainties are statistical and systematic, respectively.}
    \begin{tabular}{c|rr}
    Interval in \pt & \multicolumn{2}{c}{\dsigmadpt~{[pb/GeV]}} \\
    {[GeV]} & \multicolumn{1}{c}{\Wpmn} & \multicolumn{1}{c}{\Wmmn} \\
    \hline 
28--30 & $11.93 \pm 0.44 \pm 0.36$ & $14.81 \pm 0.47 \pm 0.35$ \\
30--32 & $14.36 \pm 0.46 \pm 0.29$ & $15.70 \pm 0.48 \pm 0.26$ \\
32--34 & $17.66 \pm 0.48 \pm 0.31$ & $15.59 \pm 0.48 \pm 0.24$ \\
34--36 & $18.87 \pm 0.51 \pm 0.31$ & $16.09 \pm 0.48 \pm 0.29$ \\
36--38 & $22.73 \pm 0.56 \pm 0.36$ & $16.53 \pm 0.49 \pm 0.24$ \\
38--40 & $23.50 \pm 0.58 \pm 0.31$ & $14.57 \pm 0.48 \pm 0.32$ \\
40--42 & $17.16 \pm 0.53 \pm 0.31$ & $10.26 \pm 0.42 \pm 0.24$ \\
42--44 & $10.45 \pm 0.43 \pm 0.30$ & $6.13 \pm 0.35 \pm 0.18$ \\
44--46 & $6.01 \pm 0.35 \pm 0.17$ & $3.28 \pm 0.28 \pm 0.23$ \\
46--48 & $3.46 \pm 0.30 \pm 0.14$ & $2.41 \pm 0.24 \pm 0.14$ \\
48--50 & $2.59 \pm 0.26 \pm 0.13$ & $1.65 \pm 0.22 \pm 0.15$ \\
50--52 & $1.75 \pm 0.21 \pm 0.14$ & $1.42 \pm 0.18 \pm 0.11$\\
    \end{tabular}
    \label{tab:combined_xs_results}
\end{table}

\subsection{Systematic uncertainties}\label{sec:systematics_dxs}

Systematic uncertainties are obtained by rerunning the differential cross-section fit with each systematic variation applied and comparing the new results to the baseline values.
The total systematic uncertainty on the differential cross-section measurement is represented by a single $24\times 24$ covariance matrix. This corresponds to twelve \pt intervals for \Wp and \Wm bosons respectively, because some sources of uncertainty are strongly correlated (or anticorrelated) between the two.
A covariance matrix is defined for each source and these are added linearly, and for some sources the covariances for more than one subsource are combined.
The sources of systematic uncertainty are categorised as follows.
\begin{description}
      
    \item \textbf{Muon efficiency}: The statistical uncertainties associated with the trigger, tracking and identification efficiency of muons are propagated to the differential cross-section fit by repeatedly varying each correction factor according to a Gaussian distribution defined by its central value and associated uncertainty, and refitting the cross-section. The effect of variations in the $\eta$ binning schemes is found to be negligible compared to the other uncertainties in this analysis.
  
    \item \textbf{Charge-independent momentum biases}: The covariance of the momentum smearing fit result is propagated through the simulation by repeating the cross-section fit with random sampling.

   \item \textbf{Charge-dependent momentum biases}: The statistical uncertainties on the pseudomass corrections are uncorrelated between the $(\eta, \phi)$ intervals. While the corrections are applied to the data, their statistical uncertainties are propagated by applying charge-dependent curvature shifts to the muon momenta in the simulation. The shifts are sampled from normal distributions representing the corrections and their uncertainties. The differential cross-section fit is repeated several times and the standard deviation in the fit results is defined as the uncertainty.

    \item \textbf{Hadronic background}: The corresponding uncertainties are fourfold. The first is the covariance from the fits of the momentum-dependent misidentification function propagated by random sampling. The second uncertainty is determined by varying the weights assigned separately to pions, kaons, and protons by a conservative $\pm 20\%$~\cite{LHCb-PAPER-2011-037}. The third uncertainty is the statistical uncertainty associated with the isolation calibration factor. The fourth uncertainty considers a different model that includes the hadronic background \pt shape, which is described in Eq.~\ref{eq:qcd_bkg_pt_model}. This alternative model (weight strategy) includes one additional factor,
\begin{equation}
\left( 1 + w \frac{\pt - \bar{p}_{\rm T}}{N_{\text{bins}}} \right),
\label{eq:qcd_bkg_pt_model}
\end{equation}
where $w$ modifies the shape of the \pt distribution, \pt is the bin centre in muon transverse momentum, and $\bar{p}_{\rm T}$ is the median \pt value (40\gev). 
The default fit model corresponds to $w=0$.

    \item \textbf{Isolation}: This includes the statistical uncertainty of the isolation calibration factors and the choice of the number of isolation intervals, which is changed from eight to six or ten. The statistical uncertainty on the isolation calibration factor is also propagated by repeating the cross-section fit with this factor shifted up and down by one standard deviation.

    \item \textbf{Unfolding}: This accounts for the uncertainty resulting from increasing or decreasing the counts of underflow and overflow intervals by 10\%. Furthermore, the number of reconstructed \pt bins is varied, increasing from twelve to twenty-four while keeping twelve true \pt bins.
     
\end{description}

\begin{figure}\centering 
    \includegraphics[width=\DefaultFigureWidth]{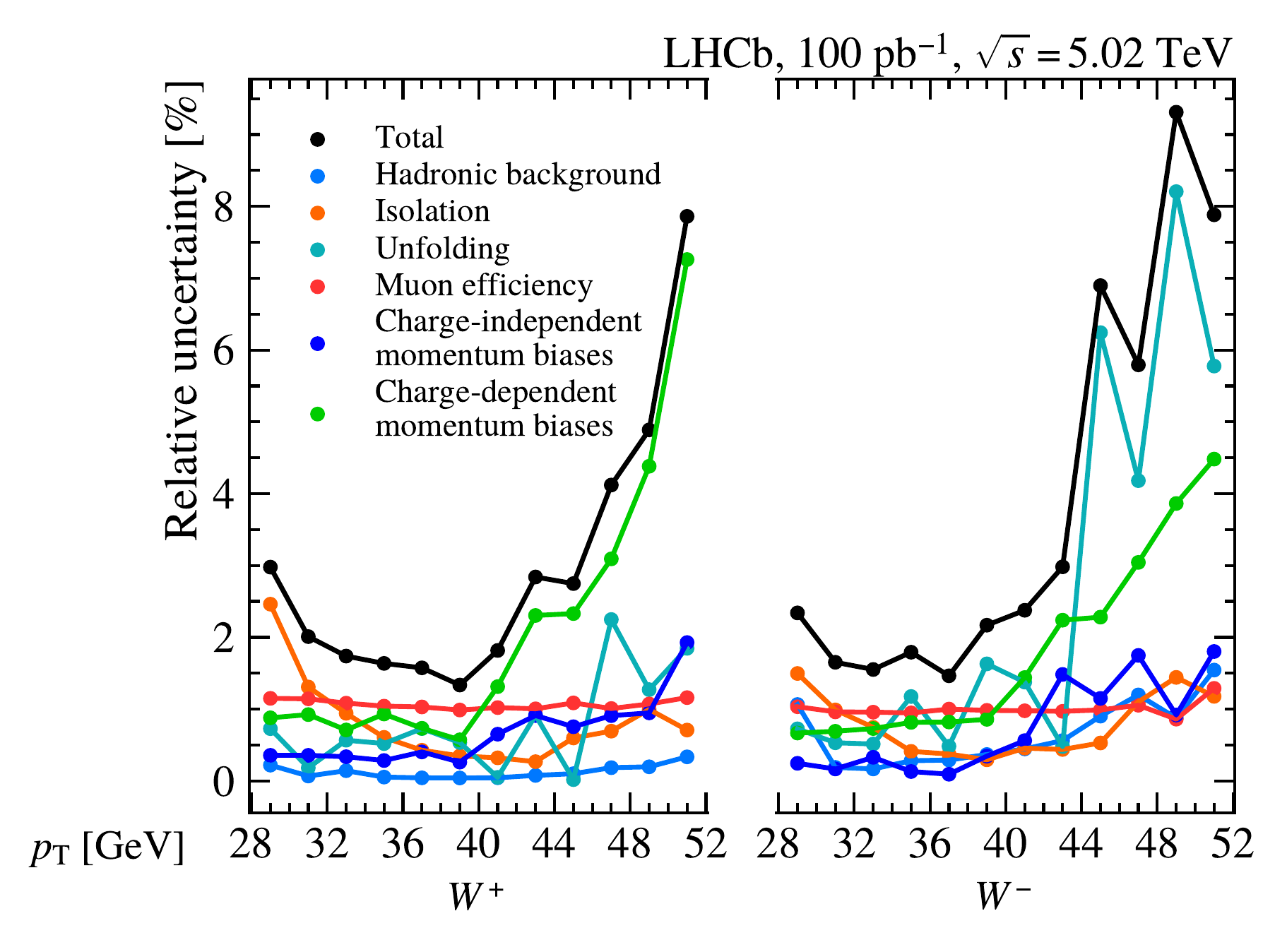}
    \caption{Relative systematic uncertainties on the (left) $W^+$ and (right) $W^-$ differential cross-sections as a function of $p_\mathrm{T}$. These are obtained by dividing the systematic uncertainty in each bin by the corresponding cross-section value, as listed in Table~\ref{tab:combined_xs_results}.}
    \label{fig:fit_systematic_relative_uncertainties_subcategories}
\end{figure}

\begin{figure}[tbp]\centering
    \includegraphics[width=\DefaultFigureWidth]{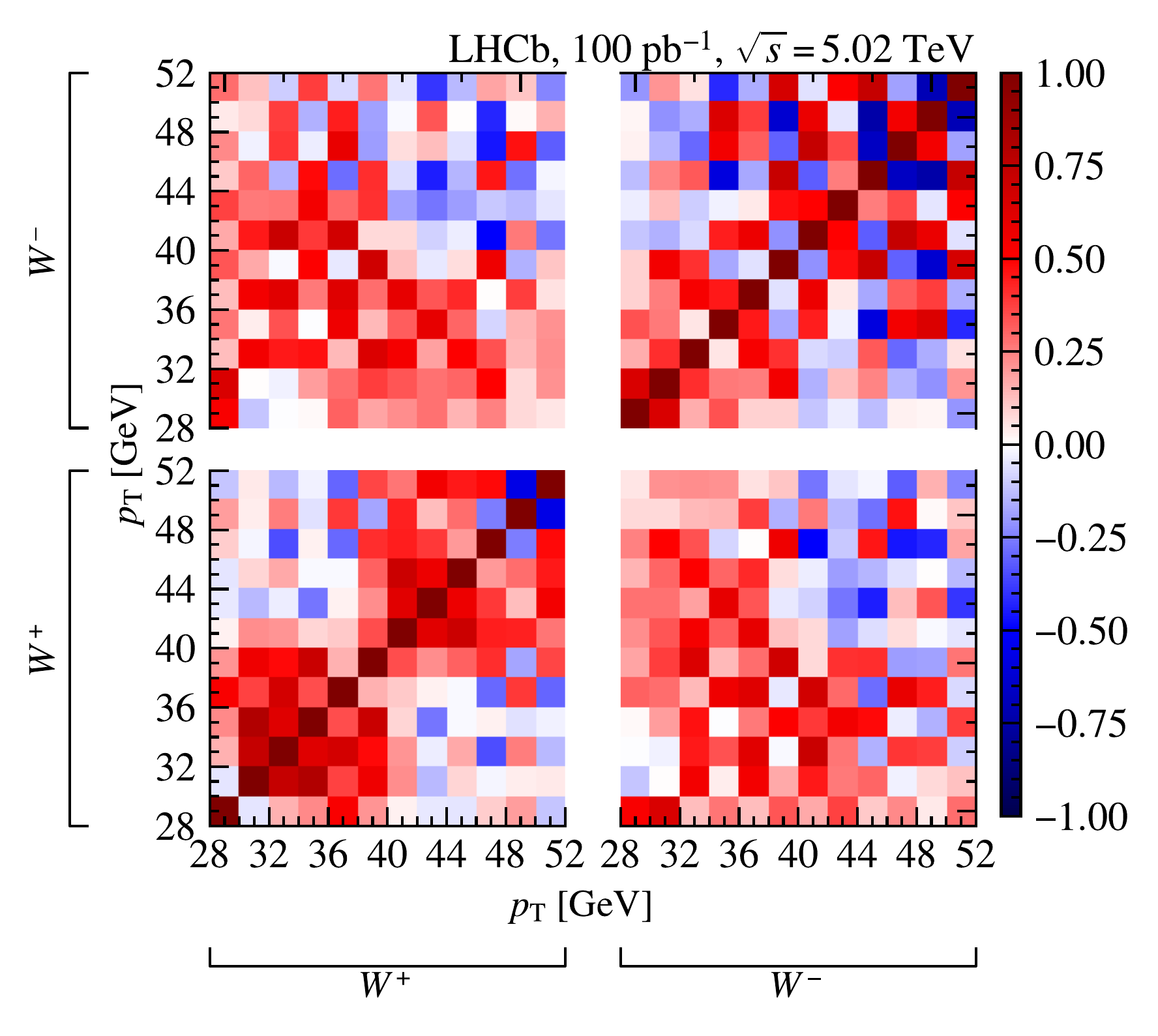}
    \caption{Correlation matrix corresponding to the total systematic uncertainty, with rows and columns ordered as $W^+$ and then $W^-$.}
    \label{fig:fit_systematic_correlation_matrices}
\end{figure}

Figure~\ref{fig:fit_systematic_relative_uncertainties_subcategories} shows the relative uncertainty for each category.  Figure~\ref{fig:fit_systematic_correlation_matrices} shows the correlation matrix corresponding to the sum of the covariance matrices for each category.
The largest source of uncertainty for the \Wp boson is the charge-dependent momentum bias, which reaches around 7\% in the highest \pt interval.  
However, this source is strongly anticorrelated between the \Wp and \Wm bosons, which is the cause of the regions of negative correlation seen in Fig.~\ref{fig:fit_systematic_correlation_matrices}. 
It is crucial that this anticorrelation is correctly accounted for in a fit for \mw. The largest source of uncertainty for the \Wm boson is the so-called unfolding uncertainty, which peaks at around 8\% at higher \pt intervals. This is driven by varying the number of reconstructed-level \pt bins from twelve to twenty-four.
The next largest source of uncertainty for both charges is the muon efficiency, at a level of around 1\%. This source is strongly correlated across all twenty-four \pt intervals. 

\section{Integrated cross-sections and comparison with predictions}
\label{sec:integrated_cross_sections}
\begin{figure}[!b!]\centering
    \includegraphics[width=\DefaultFigureWidth]{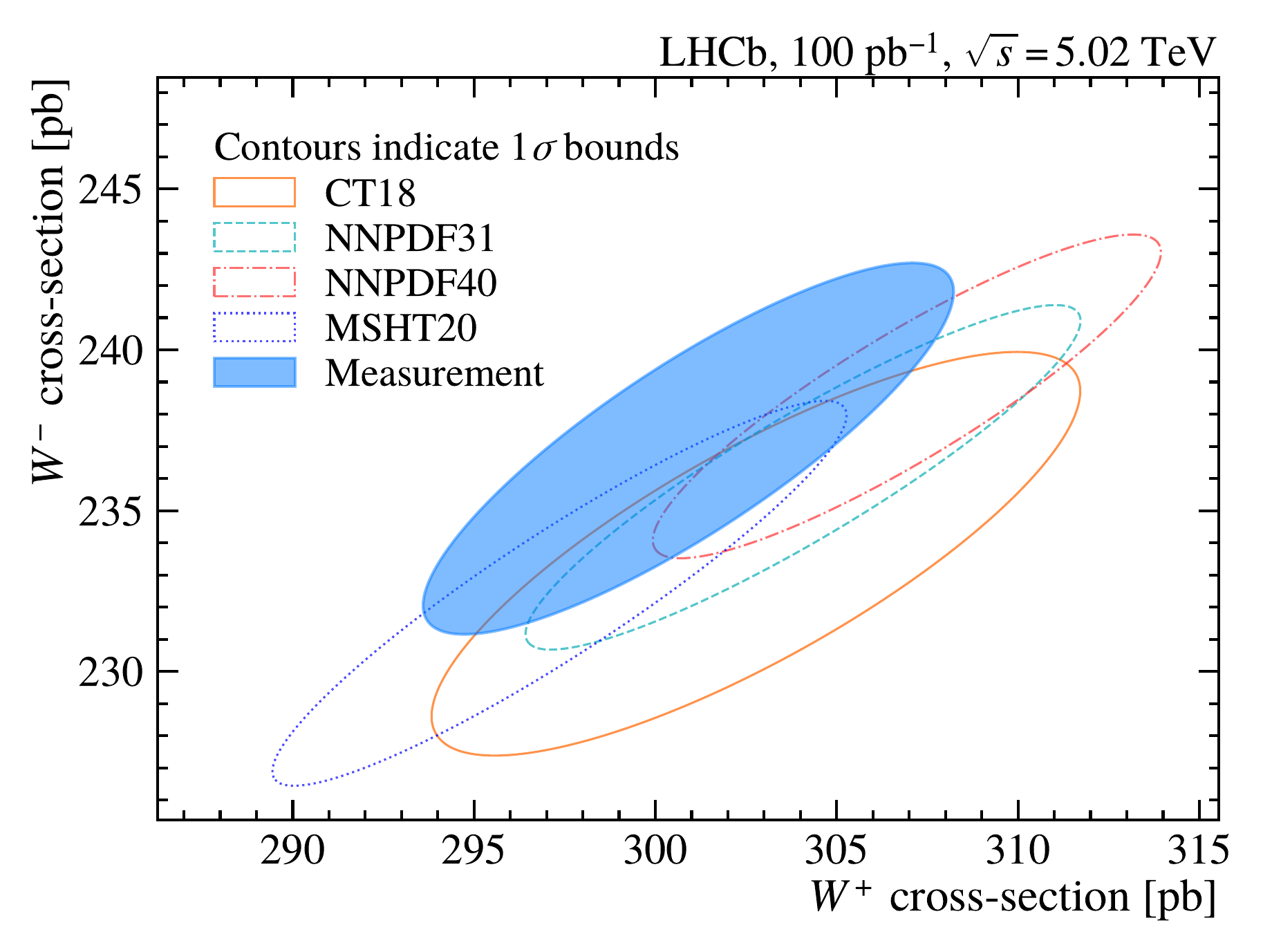}
    \caption{Integrated cross-sections compared to $\mathcal{O}(\alpha_s^2)$ predictions.}
    \label{fig:ixs_comparison_2D}
\end{figure}

The total fiducial cross-section values are computed by integrating the unfolded differential cross-sections obtained during the background subtraction fit, and the uncertainties are propagated accordingly using the information from the covariance matrices.
This integration yields:
\begin{align*}
     \sigma_\Wpmn &= \IntegratedXsecWpValue \pm \IntegratedXsecWpStatError \pm \IntegratedXsecWpSystError \pm \IntegratedXsecWpLumiError \pb, \\
    \sigma_\Wmmn &= \IntegratedXsecWmValue \pm \IntegratedXsecWmStatError \pm \IntegratedXsecWmSystError \pm \IntegratedXsecWmLumiError \pb,
\end{align*}
where the first uncertainty is statistical, the second is systematic, and the third comes from the integrated luminosity.
The correlation between the total uncertainties in the \Wp and \Wm cross-sections is \IntegratedXsecTotalErrorCorrelationWpWm. 
Figure~\ref{fig:ixs_comparison_2D} compares these results with predictions at $\mathcal{O}(\alpha_s^2)$ from \mcfm with the CT18~\cite{Hou:2019efy}, NNPDF3.1~\cite{Ball:2017nwa}, NNPDF4.0~\cite{NNPDF:2021njg} and MSHT20~\cite{Bailey:2020ooq} PDF sets.\footnote{These PDFs are evaluated at NNLO.} The predictions obtained with the \mcfm program correspond to the Born level. To approximate the effect of QED final-state radiation, \pythia Drell--Yan samples were used to compute the overall ratio of events in the fiducial region at Born level to those at bare level. This ratio is then applied to correct the MCFM predictions, so that the final integrated cross-section predictions correspond to the bare level.

The presented predictions include uncertainties from the PDF sets and variations in the scales. The renormalisation and factorisation scales are independently multiplied by factors of 0.5 and 2 around their central values, which correspond to the \PW-boson mass. Of the $3\times 3$ possible combinations, those in which the two scales vary in opposite directions are excluded.
The associated uncertainty is defined as the maximum variation out of the seven remaining scale combinations~\cite{Hamilton:2012np}.
The measurement is consistent with the predictions.

\section{\boldmath Determination of the $W$-boson mass}
\label{sec:mw_fit}

\subsection{Overview of the fit model}

The differential cross-section data can now be used in a fit for \mw.
The fit minimises the $\chisq$ between the data and theoretical templates with \mw as a free parameter.
The model is derived from that used in the \mw fit of Ref.~\cite{LHCb-PAPER-2021-024}.
The \mbox{$pp \to W^{\pm} (\to \mu^{\pm} \nu_{\mu}) X
$} events are generated using \pythia with the same version and tune as the samples with full detector simulation. 
As in Ref.~\cite{LHCb-PAPER-2021-024}, the model can mimic a variation in \mw with the use of weights from an analytic function of the true boson mass. 
Weights are assigned to map these samples to the average of the predictions of \pythia, \photos and \herwig~\cite{Bellm:2015jjp}, in the distribution of the energy difference between the dilepton system before and after final-state radiation. 

Since \pythia only includes the leading \pt-logarithms in the perturbative series in the strong coupling $\alpha_s$, through its parton shower, further corrections are required.
Following the approach of Ref.~\cite{LHCb-PAPER-2021-024}, the differential cross-section is factorised into an unpolarised factor and a closed-form sum of angular functions with eight angular coefficients. These are differential in the dilepton rapidity and \pt.\footnote{These also have a mass dependence, but this has a negligible effect in the present analysis.}
The two angular degrees of freedom are defined in the Collins--Soper frame~\cite{CollinsSoperFrame1977}.
The event sample is assigned weights based on the ratio of the cross-sections of \pythia compared to a target prediction.

The target predictions are based on the \dyturbo~\cite{DYTurbo2020} program.
Following Refs.~\cite{LHCb-PAPER-2021-024, LHCb-PAPER-2024-028}, the analysis is performed with models based on the NNPDF3.1, MSHT20, and CT18 PDF sets, 
and the final result is a simple arithmetic average of those obtained with the three separate PDF sets.
The unpolarised cross-sections are based on the summation of the large $\pt$-logarithms up to the second order, followed by matching to exact $\mathcal{O}(\alpha_s^2)$ calculations. 
The angular coefficients are calculated at $\mathcal{O}(\alpha_s)$ accuracy.
For the unpolarised cross-sections, the predictions 
are made for a range of $\alpha_s$ and $g$ values, where $g$ encodes the nonperturbative dynamics affecting the $\pt$ distribution of the partons.

\subsection{\boldmath Validation fit to $Z$-boson production cross-section data}\label{subsec:z_fit}

\begin{figure}[tbp]
    \centering
    \includegraphics[width=\DefaultFigureWidth]{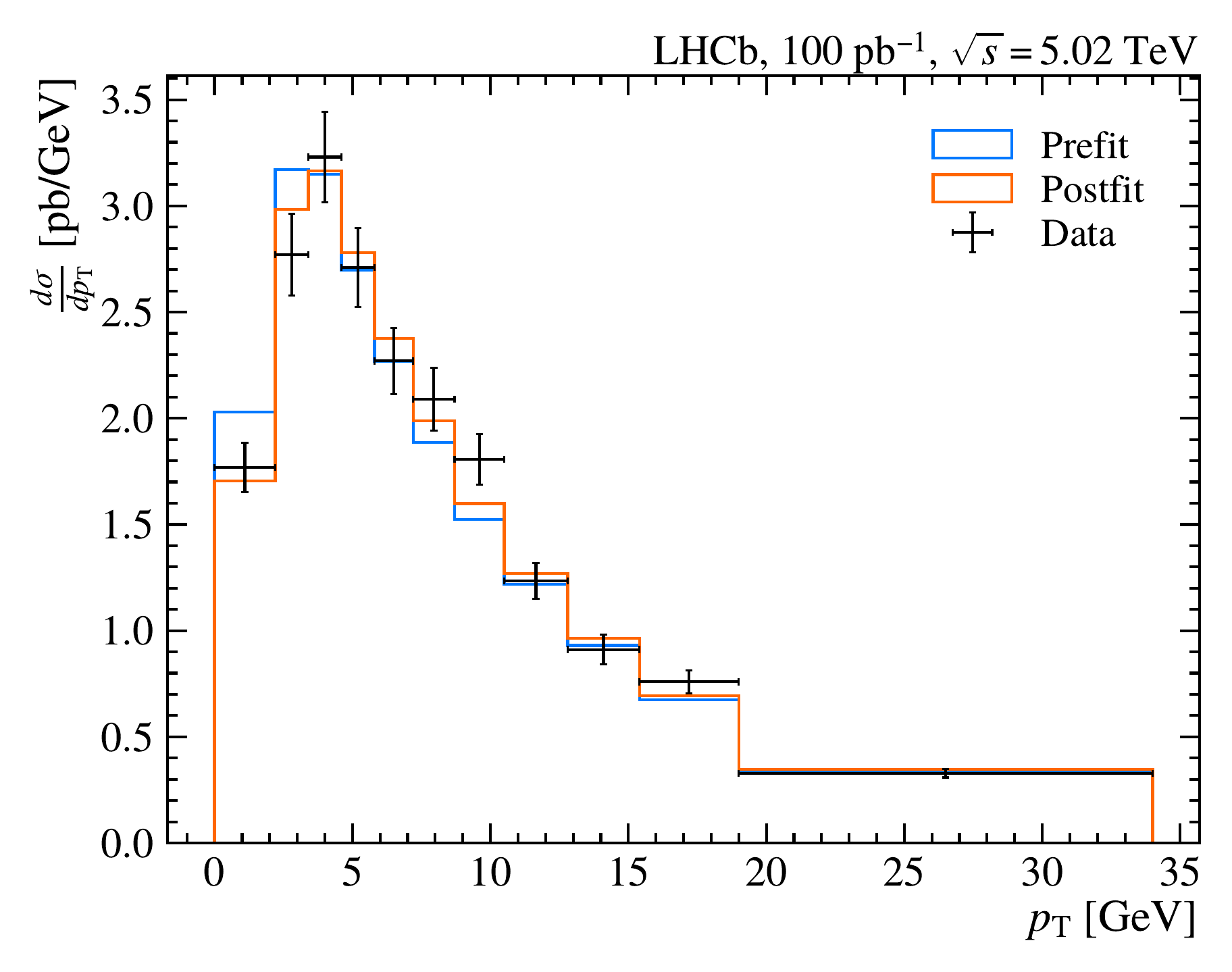}
    \caption{Differential cross-section data for $Z$-boson production from Ref.~\cite{LHCb-PAPER-2023-010} compared to the physics model to be used in the $\ensuremath{m_W}$ determination, before and after fitting for $\alpha_s$ and $g$.}
    \label{fig:Z_fit}
\end{figure}

Figure~\ref{fig:Z_fit} shows the $Z$-boson $\deriv\sigma/\deriv\pt$ data at $5.02\tev$ published by the LHCb collaboration~\cite{LHCb-PAPER-2023-010}.
The data are compared to the model described in the previous section, with the NNPDF3.1 PDF set and with $\alpha_s=0.118$ and $g=1\gev^2$. 
It can be seen that the \dyturbo prediction, after normalising to the data, overestimates the cross-section in the first two bins, but is below the data in a broad region around $10\gev$.
A fit in which the normalisation, $\alpha_s$ and $g$ are allowed to vary freely gives a $\chisqndf$ of $\minimumZfit$.
Furthermore, it favours an increased value of $g=\gZfit\pm \gZfitError \gevgev$, which suppresses the first two bins, and a decreased value of $\alpha_s=\asZfit\pm \asZfitError$,\footnote{This should not be considered as a robust determination of $\alpha_s$. A full assessment of the theoretical uncertainties is beyond the scope of the present analysis.} which improves the description at higher \pt. The preference for a smaller \as value observed in this analysis, compared to the PDG value~\cite{PDG2024}, is similar to what has been reported in other LHCb analyses, including the \PW mass measurement~\cite{LHCb-PAPER-2021-024}. This recurring trend across resummed predictions of the Drell–Yan \pt distribution at the LHC highlights an interesting open question. Further theoretical work will be valuable to understand its origin. 

\subsection{\boldmath Determination of the $W$-boson mass}

 \begin{figure}[tbp]
    \centering
    \includegraphics[width=\DefaultFigureWidth]{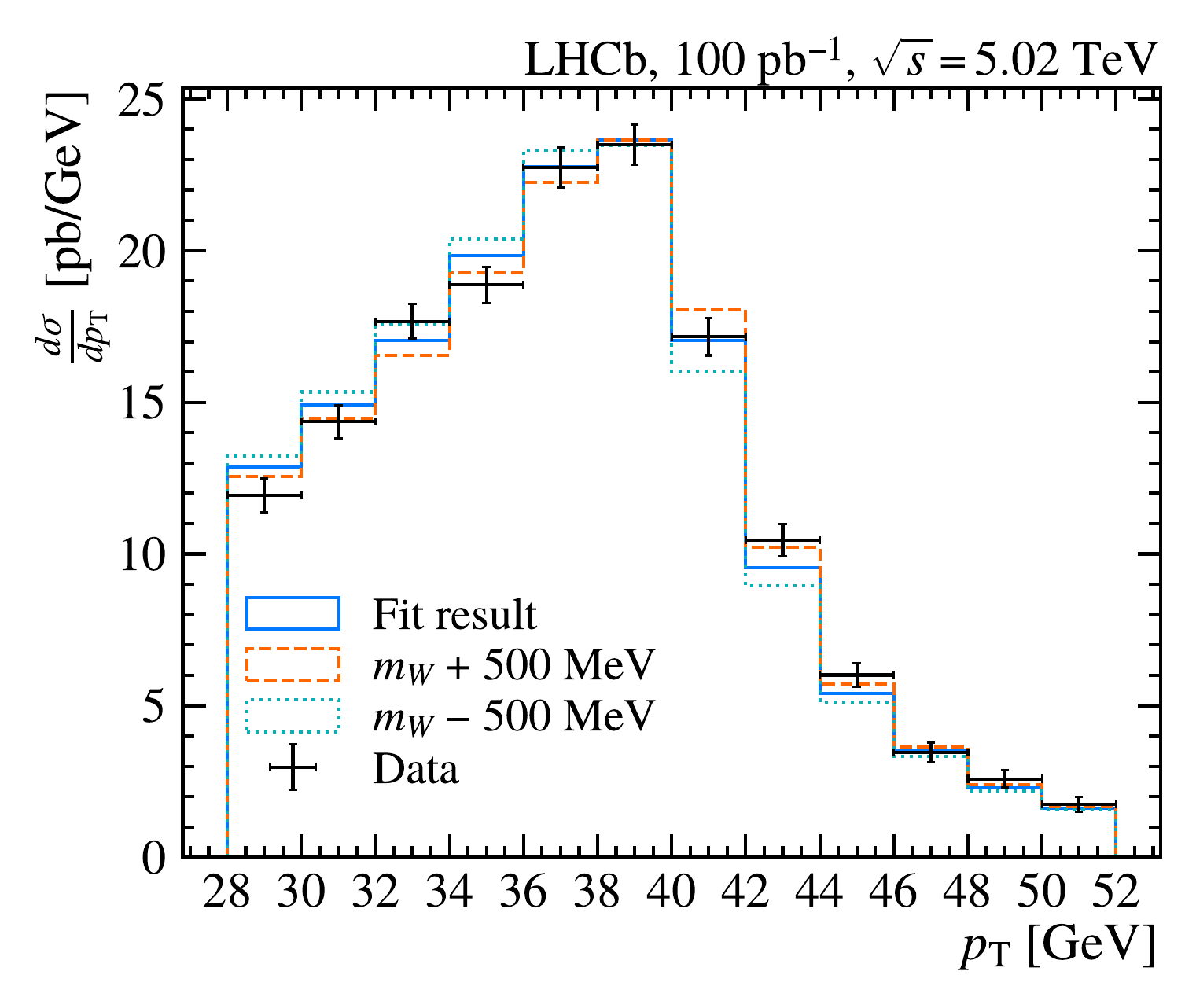}
    \includegraphics[width=\DefaultFigureWidth]{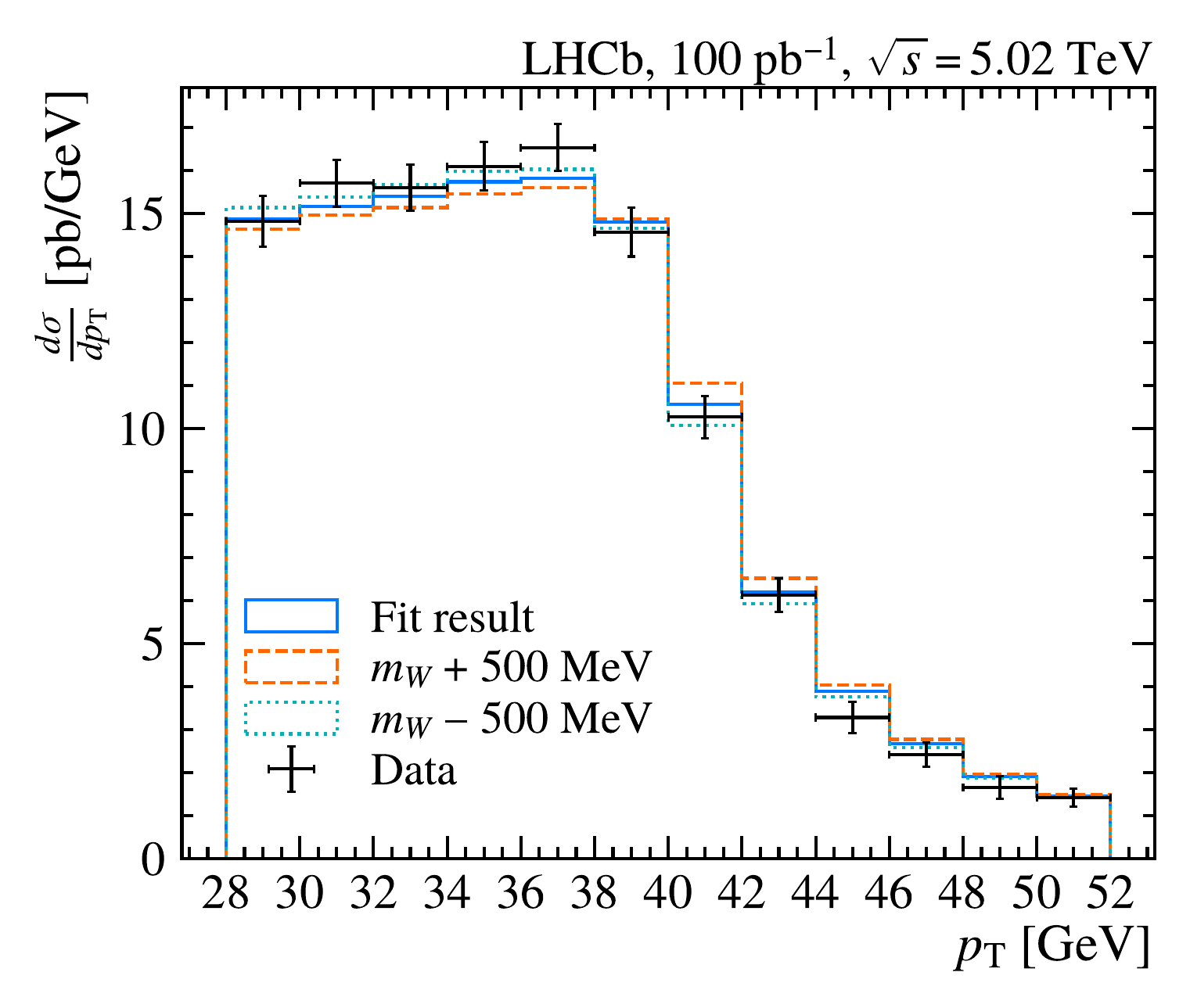}
    \hfill
    \caption{Comparison of the (upper) $W^+$ and (lower) $W^-$ ${\ensuremath{\text{d}\sigma/\text{d}p_\mathrm{T}}}$ data with the results of the $\ensuremath{m_W}$ fit. The $\ensuremath{m_W}$ fit results shifted by two extreme values of $\pm 500 \text{\,Me\kern-0.1em V}$ are also shown for illustration.}
    \label{fig:default_mW_with_syst}
\end{figure} 

Figure~\ref{fig:default_mW_with_syst} shows the \PW-boson \dsigmadpt data compared to the results of a fit, based on the NNPDF3.1 PDF set, with the $\alpha_s$ value fixed to that determined from the \PZ-boson data using the same PDF set, and the $g$ and \mw values varying freely.
This gives a \chisqndf of $\mwFitMinChisq/21$, and best-fit values of $m_W = \mwFitValue \pm \mwFitExperimentError \mev$ and $g = \gValueWithSystematics \pm \gErrorWithSystematics \gevgev$.
This fit includes all sources of uncertainty on the \dsigmadpt measurement, via the $24\times 24$ covariance matrix. 
The following additional sources of theoretical uncertainties are assessed for the mass measurement:
\begin{description}
    \item {\bf Strong coupling value}: The statistical uncertainty from the fit to the \PZ-boson data is propagated by repeating the \mw fit with this value shifted by one standard deviation in the two directions. The resulting uncertainty is $\StrongCouplingError \mev$.
    \item {\bf Perturbative accuracy in the strong coupling}: The error due to missing higher-order terms in the strong coupling $\alpha_s$ is assessed by varying the renormalisation and factorisation scales using the same prescription described for the \mcfm predictions in Sec.~\ref{sec:integrated_cross_sections}. The resulting uncertainty is $\qcdError \mev$.
    \item {\bf QED accuracy}: As in Ref.~\cite{LHCb-PAPER-2021-024}, three additional \mw fits are performed. Each uses one of the three different models already described, rather than their average. The largest variation of $\qedError \mev$ is assigned as the uncertainty.
    \item {\bf Parton distribution functions}: Each of the three used PDF sets is provided with a prescription for assessing the uncertainty on an observable. 
    For the NNPDF3.1 set this is based on taking the RMS of predictions computed with 100 different ``replica" PDF sets.
    Eigenvectors are used for the uncertainties on the other two PDF sets.
    Assuming full correlation between the three PDF sets, the PDF uncertainty on the arithmetic average of the three is $\avgPdfReplicaError$\mev.
\end{description}

\begin{table}\centering 
\caption{\label{tab:other_pdfs}Results of the \mw fit with the NNPDF3.1, MSHT20 and CT18 PDF sets.
The \mw fit result includes the total covariance of the differential cross-section measurement.
The theory uncertainty $\sigma_{\rm theory}$ includes all other sources except for the PDFs themselves.}
\begin{tabular}{lcccc}
PDF set & $\chisqndf$ & \mw & $\sigma_{\rm theory}$ & $\sigma_{\rm PDF}$ \\
\hline
NNPDF3.1 & 25.6/21 & $\mwFitValue \pm \mwFitExperimentError \mev$ & $25\mev$ & $28\mev$ \\
MSHT20 & 27.2/21 & $80380 \pm 129\mev$ & $25\mev$ & $15\mev$\\
CT18 & 23.9/21 & $80362 \pm 130\mev$ & $25\mev$ & $23\mev$ \\
\end{tabular}
\end{table}

Table~\ref{tab:other_pdfs} lists the \mw fit results for the three different PDF sets. 
The variation in the three central values is smaller than, but comparable in size to, the PDF uncertainties for the three sets.
The final quoted value of \mw is the average of the three values obtained by using the three sets of PDFs. The PDF uncertainty added to the overall theoretical uncertainty is calculated as an average of individual uncertainties for each set of PDFs. The final result is
\begin{equation*}
\mw = \AvgMwFitValueWithSyst \pm \mwFitExperimentError \pm \AvgMwFitTheoryError \mev,
\end{equation*}
where the first uncertainty reflects experimental contributions, primarily from the measurement of $\deriv\sigma/\deriv\pt$, and the second represents theoretical uncertainties. The latter is obtained by combining, in quadrature, the QCD and QED uncertainties together with the averaged PDF uncertainty.
This result is compatible with earlier experimental measurements (see Sec.~\ref{sec:Introduction}) and with the global electroweak fit~\cite{Gfitter}.

\begin{table}
\caption{\label{tab:exp_syst_impacts_on_mw}Impact of each source of experimental systematic uncertainty on the $m_W$ uncertainty. The impact is defined as the square root of the difference in quadrature between the statistical uncertainty and the uncertainty including a single systematic source. The changes in the fit $\chisq$ and best-fit $m_W$ values are also shown.  The shift in the \PW-boson mass, denoted $\Delta\mw$, is defined as the value after the change minus the value before. Impacts or changes in the $\chisq$ of less than one unit are indicated with a dash.}
\begin{tabular}{l|c|cc}
Source & Impact [MeV] & $\Delta\chi^2$ & $\Delta m_W$ [MeV]\\
\hline 
Charge-independent momentum biases & 58 & -- & $-8$\\
Charge-dependent momentum biases & 41 & $-19$ & $-150$ \\
Hadronic background & 7 & $-2$ & $+13$ \\
Muon efficiency & 15 & $-1$ & $+8$ \\
Isolation & 15 & -- & $-3$ \\
Unfolding & -- &  $-1$ & $+7$\\
\hline 
Quadrature sum & 75 &  &  \\
\end{tabular}
\end{table}

All sources of uncertainty on the differential cross-section measurement are already accounted for in the ``experimental"
uncertainty of $\mwFitExperimentError\mev$, through the total covariance of the measurement.
However, it is informative to decompose the impact of the different sources of systematic uncertainty on the differential measurement.
Table~\ref{tab:exp_syst_impacts_on_mw} lists, for each source of systematic uncertainty, the impact on the precision of the \mw fit result,
defined  as the quadrature difference in uncertainty when that source is included on its own, compared with the case where no uncertainty sources are included.
The charge-independent momentum bias uncertainty has the largest impact, of $\ChargeIndependentMomentumBiasImpact \mev$.
The next most important source is the charge-dependent momentum biases. When this uncertainty is removed from the fit, there is also a large shift in the best-fit \mw value; however, there is a corresponding large increase in $\chisq$, so the data have insufficient error coverage when this source is not included. 
The other two categories of uncertainty have impacts of $15\mev$ or less.
 The statistical uncertainty on the measurement can be estimated by subtracting, in quadrature, the total uncertainty in Table~\ref{tab:exp_syst_impacts_on_mw} from the experimental uncertainty coming from the background subtraction fit, and amounts to about $\mwStatOnlyErrorQuadrature\mev$.

\section{Cross-checks}\label{sec:cross-checks}

The first set of cross-checks looks at the sensitivity of the measured \mw value to changes in the differential cross-section measurement.
\begin{description}
\item \textbf{Sensitivity to the \boldmath{\mw} value in the simulation}: The value of \mw in the signal simulation used to construct the response matrix is shifted by $+100\mev$. This results in a shift of only $\mWCrossCheckImpact\mev$  in the determination of \mw.
\item \textbf{The \boldmath{\PW} production model}: The baseline result uses the \pythia description of the \PW-boson production. If the response matrix is constructed with the \pythia events weighted to match the unpolarised cross-section predictions of \dyturbo, the final value of \mw is shifted by $\UnpolarisedXSCrossCheckImpact\mev$.
If, instead, the angular distributions are weighted based on the angular coefficients of \dyturbo, the value of \mw is shifted by $\PolarisedXSCrossCheckImpact\mev$.
\end{description}

\noindent The second set of cross-checks considers variations only in the \mw fit.
\begin{description}
\item \textbf{Fit parameters:} By default, the \mw fit uses the $\alpha_s$ value determined from the fit to the \PZ-boson data, but allows $g$ to vary freely to a value preferred by the \PW-boson data. 
If the $g$ value is also fixed to the larger value found in the fit to the \PZ-boson data, \mw shifts by $\FixGCrossCheckImpact\mev$. If, instead, $\alpha_s$ is allowed to vary independently for the \mw fit, the shift in \mw is of $\FloatAlphaCrossCheckImpact\mev$. 
\item \textbf{QCD scales:} The baseline \dyturbo predictions are made with the renormalisation and factorisation scales parametrised as the quadrature sum of the dilepton mass and \pt.  
An alternative scheme, with the scales set to the dilepton mass, results in a shift in \mw of $\qcdScalesCrossCheckImpact\mev$.
\item \textbf{QCD cut-off treatment:} \dyturbo uses an inverse Bessel transform and has a default cut-off value of $3\gev$ that avoids the Landau pole. Changing the cut-off value to $1\gev$, the \mw value shifts by only $\qcdCutOffCrossCheckImpact\mev$.
\item \textbf{Partial inclusion of higher orders:} A \dyturbo prediction at N3LL and N3LO, with default values of $\alpha_s$ and $g$, is used to transform the baseline (N2LL and N2LO) predictions. This results in a shift in \mw of $\HigherOrderCrossCheckImpact\mev$. 
\item\textbf{PDF set:} The three PDF sets used in the \mw determination are consistent with those used in previous LHCb analyses~\cite{LHCb-PAPER-2021-024, LHCb-PAPER-2024-028}.
The result with the NNPDF4.0 PDF set, which was not included, is shifted by $-5\mev$ compared to the result with the NNPDF3.1 PDF set.

\end{description}

In summary, the variations observed in all cross-checks are minor in relation to experimental and theoretical uncertainties and align with the baseline result. This agreement provides additional confidence in the robustness and reliability of the measurement.

\section{Conclusion}
\label{sec:conclusion}

This paper reports a measurement of the differential cross-section for \mbox{$pp \to W^{\pm} (\to \mu^{\pm} \nu_{\mu}) X$} production in muon \pt intervals.
For the first time, we demonstrate the feasibility of using such differential data in a subsequent determination of the \PW-boson mass.

The cross-sections are measured at a $pp$ centre-of-mass energy of \mbox{$\sqrt{s}= 5.02$\tev} using a limited dataset corresponding to an integrated luminosity of \mbox{$\IntegratedLumiValue\invpb$}, recorded during 2017 by the LHCb experiment.
  Integrated over the muon pseudorapidity range $2.2 < \eta < 4.4$, the cross-sections are measured differentially in twelve intervals of muon transverse momentum in the range \mbox{$28 < \pt < 52\gev$}.
  The measured cross-sections integrated over \pt are
  \begin{align*}
    \sigma_\Wpmn &= \IntegratedXsecWpValue \pm \IntegratedXsecWpStatError \pm \IntegratedXsecWpSystError \pm \IntegratedXsecWpLumiError \pb, \\
    \sigma_\Wmmn &= \IntegratedXsecWmValue \pm \IntegratedXsecWmStatError \pm \IntegratedXsecWmSystError \pm \IntegratedXsecWmLumiError \pb,
  \end{align*}
  where the first, second and third uncertainties are statistical, systematic and due to the luminosity determination, respectively. 
  These results are consistent with theoretical predictions at fixed-order in the strong coupling.
  As a proof of principle, the differential results are used to determine the \PW-boson mass
  \begin{equation*}
    \mw = \AvgMwFitValueWithSyst \pm \mwFitExperimentError \pm \AvgMwFitTheoryError \mev,
  \end{equation*}
  where the first uncertainty is experimental and the second is theoretical.
  This result is in agreement with earlier measurements and indirect determinations based on electroweak precision data. Based on the differential cross-section data reported here, further study can now be performed on the modelling of the Drell--Yan process in the forward rapidity region covered by the LHCb experiment.
  The analysis approach can be applied to the LHCb Run~2 dataset at $\sqrt{s}=13\tev$, which contains roughly two orders of magnitude more \PW-boson decays than the small dataset analysed here, with an expected statistical uncertainty on \mw of around $12\mev$ as well as reduced systematic uncertainties. Looking ahead, it is expected that Run~3 data will provide improved precision, benefitting from increased statistics and improved detector performance.

\section*{Acknowledgements}
%
%
\noindent We express our gratitude to our colleagues in the CERN
accelerator departments for the excellent performance of the LHC. We
thank the technical and administrative staff at the LHCb
institutes.
We acknowledge support from CERN and from the national agencies:
ARC (Australia);
CAPES, CNPq, FAPERJ and FINEP (Brazil); 
MOST and NSFC (China); 
CNRS/IN2P3 (France); 
BMFTR, DFG and MPG (Germany);
INFN (Italy); 
NWO (Netherlands); 
MNiSW and NCN (Poland); 
MCID/IFA (Romania); 
MICIU and AEI (Spain);
SNSF and SER (Switzerland); 
NASU (Ukraine); 
STFC (United Kingdom); 
DOE NP and NSF (USA).
We acknowledge the computing resources that are provided by ARDC (Australia), 
CBPF (Brazil),
CERN, 
IHEP and LZU (China),
IN2P3 (France), 
KIT and DESY (Germany), 
INFN (Italy), 
SURF (Netherlands),
Polish WLCG (Poland),
IFIN-HH (Romania), 
PIC (Spain), CSCS (Switzerland), 
and GridPP (United Kingdom).
We are indebted to the communities behind the multiple open-source
software packages on which we depend.
Individual groups or members have received support from
Key Research Program of Frontier Sciences of CAS, CAS PIFI, CAS CCEPP, 
Fundamental Research Funds for the Central Universities,  and Sci.\ \& Tech.\ Program of Guangzhou (China);
Minciencias (Colombia);
EPLANET, Marie Sk\l{}odowska-Curie Actions, ERC and NextGenerationEU (European Union);
A*MIDEX, ANR, IPhU and Labex P2IO, and R\'{e}gion Auvergne-Rh\^{o}ne-Alpes (France);
Alexander-von-Humboldt Foundation (Germany);
ICSC (Italy); 
Severo Ochoa and Mar\'ia de Maeztu Units of Excellence, GVA, XuntaGal, GENCAT, InTalent-Inditex and Prog.~Atracci\'on Talento CM (Spain);
SRC (Sweden);
the Leverhulme Trust, the Royal Society and UKRI (United Kingdom).


\section*{Appendix}

\appendix

\section{Total covariance matrix}
\label{sec:total_cov_matrix}
The total covariance matrix (statistical + systematic) used to fit \mw, as demonstrated  in Sec.~\ref{sec:mw_fit}, is given in Listing~\ref{lst:cov_matrix}.

\begin{lstlisting}[basicstyle=\ttfamily\footnotesize, breaklines=true, caption={Full 24$\times$24 covariance matrix of the measurement.}, label={lst:cov_matrix}]
Row 1: 0.3192, -0.0252, 0.0287, 0.0302, 0.0696, 0.0258, 0.0048, -0.0028, -0.0022, 0.0057, 0.0089, -0.0050, 0.0644, 0.0592, 0.0113, 0.0273, 0.0114, 0.0369, 0.0145, 0.0241, 0.0085, 0.0113, 0.0022, 0.0114
Row 2: -0.0252, 0.2915, 0.0282, 0.0816, 0.0403, 0.0529, 0.0213, -0.0106, 0.0045, -0.0002, 0.0015, 0.0020, -0.0114, 0.0001, 0.0381, 0.0029, 0.0381, 0.0154, 0.0314, 0.0140, 0.0197, -0.0010, 0.0031, 0.0040
Row 3: 0.0287, 0.0282, 0.3286, 0.0106, 0.0850, 0.0457, 0.0209, -0.0029, 0.0089, -0.0147, 0.0099, -0.0054, -0.0004, -0.0021, 0.0336, 0.0299, 0.0458, -0.0015, 0.0531, 0.0150, -0.0105, 0.0169, 0.0179, -0.0033
Row 4: 0.0302, 0.0816, 0.0106, 0.3515, -0.0303, 0.0844, 0.0056, -0.0241, -0.0006, 0.0015, -0.0022, -0.0008, 0.0014, 0.0153, 0.0345, -0.0004, 0.0194, 0.0494, 0.0292, 0.0305, 0.0337, -0.0014, -0.0072, 0.0130
Row 5: 0.0696, 0.0403, 0.0850, -0.0303, 0.4428, -0.0743, 0.0321, -0.0000, 0.0002, -0.0148, 0.0177, -0.0147, 0.0379, 0.0265, 0.0118, 0.0579, 0.0541, -0.0048, 0.0592, 0.0192, -0.0229, 0.0290, 0.0242, -0.0030
Row 6: 0.0258, 0.0529, 0.0457, 0.0844, -0.0743, 0.4385, -0.0656, 0.0415, 0.0123, 0.0194, -0.0072, 0.0159, 0.0193, 0.0311, 0.0482, 0.0126, 0.0219, 0.0682, 0.0057, 0.0229, 0.0290, -0.0085, -0.0087, 0.0094
Row 7: 0.0048, 0.0213, 0.0209, 0.0056, 0.0321, -0.0656, 0.3750, -0.0223, 0.0530, 0.0155, 0.0182, 0.0112, 0.0240, 0.0266, 0.0404, 0.0286, 0.0448, 0.0119, 0.0058, -0.0103, -0.0045, 0.0030, -0.0004, -0.0018
Row 8: -0.0028, -0.0106, -0.0029, -0.0241, -0.0000, 0.0415, -0.0223, 0.2770, -0.0277, 0.0308, 0.0010, 0.0233, 0.0285, 0.0214, 0.0132, 0.0504, 0.0237, -0.0041, -0.0066, -0.0147, -0.0299, 0.0055, 0.0153, -0.0130
Row 9: -0.0022, 0.0045, 0.0089, -0.0006, 0.0002, 0.0123, 0.0530, -0.0277, 0.1482, -0.0384, 0.0187, 0.0070, 0.0084, 0.0127, 0.0200, 0.0142, 0.0166, 0.0035, -0.0013, -0.0059, -0.0054, -0.0013, 0.0001, -0.0025
Row 10: 0.0057, -0.0002, -0.0147, 0.0015, -0.0148, 0.0194, 0.0155, 0.0308, -0.0384, 0.1077, -0.0402, 0.0196, 0.0122, 0.0184, 0.0116, -0.0035, 0.0001, 0.0253, -0.0176, -0.0028, 0.0147, -0.0091, -0.0094, 0.0027
Row 11: 0.0089, 0.0015, 0.0099, -0.0022, 0.0177, -0.0072, 0.0182, 0.0010, 0.0187, -0.0402, 0.0861, -0.0373, 0.0032, 0.0024, 0.0041, 0.0054, 0.0116, -0.0060, 0.0081, -0.0031, -0.0080, 0.0082, 0.0002, 0.0016
Row 12: -0.0050, 0.0020, -0.0054, -0.0008, -0.0147, 0.0159, 0.0112, 0.0233, 0.0070, 0.0196, -0.0373, 0.0622, 0.0022, 0.0076, 0.0074, 0.0086, 0.0020, 0.0048, -0.0091, -0.0012, -0.0006, -0.0062, 0.0032, -0.0037
Row 13: 0.0644, -0.0114, -0.0004, 0.0014, 0.0379, 0.0193, 0.0240, 0.0285, 0.0084, 0.0122, 0.0032, 0.0022, 0.3450, 0.0221, 0.0260, 0.0366, 0.0104, 0.0121, -0.0080, -0.0013, -0.0097, 0.0020, 0.0013, -0.0078
Row 14: 0.0592, 0.0001, -0.0021, 0.0153, 0.0265, 0.0311, 0.0266, 0.0214, 0.0127, 0.0184, 0.0024, 0.0076, 0.0221, 0.2980, -0.0223, 0.0307, 0.0159, 0.0461, -0.0091, 0.0065, 0.0146, -0.0049, -0.0084, 0.0063
Row 15: 0.0113, 0.0381, 0.0336, 0.0345, 0.0118, 0.0482, 0.0404, 0.0132, 0.0200, 0.0116, 0.0041, 0.0074, 0.0260, -0.0223, 0.2866, -0.0498, 0.0421, 0.0300, -0.0034, -0.0039, 0.0182, -0.0098, -0.0058, 0.0017
Row 16: 0.0273, 0.0029, 0.0299, -0.0004, 0.0579, 0.0126, 0.0286, 0.0504, 0.0142, -0.0035, 0.0054, 0.0086, 0.0366, 0.0307, -0.0498, 0.3167, -0.0313, -0.0023, 0.0290, -0.0008, -0.0384, 0.0222, 0.0284, -0.0135
Row 17: 0.0114, 0.0381, 0.0458, 0.0194, 0.0541, 0.0219, 0.0448, 0.0237, 0.0166, 0.0001, 0.0116, 0.0020, 0.0104, 0.0159, 0.0421, -0.0313, 0.2996, -0.0709, 0.0469, -0.0004, -0.0087, 0.0107, 0.0142, -0.0042
Row 18: 0.0369, 0.0154, -0.0015, 0.0494, -0.0048, 0.0682, 0.0119, -0.0041, 0.0035, 0.0253, -0.0060, 0.0048, 0.0121, 0.0461, 0.0300, -0.0023, -0.0709, 0.3263, -0.0814, 0.0406, 0.0487, -0.0131, -0.0306, 0.0233
Row 19: 0.0145, 0.0314, 0.0531, 0.0292, 0.0592, 0.0057, 0.0058, -0.0066, -0.0013, -0.0176, 0.0081, -0.0091, -0.0080, -0.0091, -0.0034, 0.0290, 0.0469, -0.0814, 0.2389, -0.0289, -0.0062, 0.0223, 0.0220, -0.0016
Row 20: 0.0241, 0.0140, 0.0150, 0.0305, 0.0192, 0.0229, -0.0103, -0.0147, -0.0059, -0.0028, -0.0031, -0.0012, -0.0013, 0.0065, -0.0039, -0.0008, -0.0004, 0.0406, -0.0289, 0.1525, -0.0253, 0.0181, -0.0035, 0.0109
Row 21: 0.0085, 0.0197, -0.0105, 0.0337, -0.0229, 0.0290, -0.0045, -0.0299, -0.0054, 0.0147, -0.0080, -0.0006, -0.0097, 0.0146, 0.0182, -0.0384, -0.0087, 0.0487, -0.0062, -0.0253, 0.1302, -0.0491, -0.0178, 0.0162
Row 22: 0.0113, -0.0010, 0.0169, -0.0014, 0.0290, -0.0085, 0.0030, 0.0055, -0.0013, -0.0091, 0.0082, -0.0062, 0.0020, -0.0049, -0.0098, 0.0222, 0.0107, -0.0131, 0.0223, 0.0181, -0.0491, 0.0790, -0.0109, 0.0035
Row 23: 0.0022, 0.0031, 0.0179, -0.0072, 0.0242, -0.0087, -0.0004, 0.0153, 0.0001, -0.0094, 0.0002, 0.0032, 0.0013, -0.0084, -0.0058, 0.0284, 0.0142, -0.0306, 0.0220, -0.0035, -0.0178, -0.0109, 0.0702, -0.0306
Row 24: 0.0114, 0.0040, -0.0033, 0.0130, -0.0030, 0.0094, -0.0018, -0.0130, -0.0025, 0.0027, 0.0016, -0.0037, -0.0078, 0.0063, 0.0017, -0.0135, -0.0042, 0.0233, -0.0016, 0.0109, 0.0162, 0.0035, -0.0306, 0.0451
\end{lstlisting}

\clearpage 

\addcontentsline{toc}{section}{References}
\bibliographystyle{LHCb}
\bibliography{main,standard,LHCb-PAPER,LHCb-CONF,LHCb-DP,LHCb-TDR}

\newpage
\centerline
{\large\bf LHCb collaboration}
\begin
{flushleft}
\small
R.~Aaij$^{38}$\lhcborcid{0000-0003-0533-1952},
A.S.W.~Abdelmotteleb$^{57}$\lhcborcid{0000-0001-7905-0542},
C.~Abellan~Beteta$^{51}$\lhcborcid{0009-0009-0869-6798},
F.~Abudin{\'e}n$^{57}$\lhcborcid{0000-0002-6737-3528},
T.~Ackernley$^{61}$\lhcborcid{0000-0002-5951-3498},
A. A. ~Adefisoye$^{69}$\lhcborcid{0000-0003-2448-1550},
B.~Adeva$^{47}$\lhcborcid{0000-0001-9756-3712},
M.~Adinolfi$^{55}$\lhcborcid{0000-0002-1326-1264},
P.~Adlarson$^{85}$\lhcborcid{0000-0001-6280-3851},
C.~Agapopoulou$^{14}$\lhcborcid{0000-0002-2368-0147}
% ...

{\footnotesize \it

% ...
$^{14}$Universit{\'e} Paris-Saclay, CNRS/IN2P3, IJCLab, Orsay, France\\
% ...
$^{38}$Nikhef National Institute for Subatomic Physics, Amsterdam, Netherlands\\
% ...
$^{47}$Instituto Galego de F{\'\i}sica de Altas Enerx{\'\i}as (IGFAE), Universidade de Santiago de Compostela, Santiago de Compostela, Spain\\
% ...
$^{51}$Physik-Institut, Universit{\"a}t Z{\"u}rich, Z{\"u}rich, Switzerland\\
% ...
$^{55}$H.H. Wills Physics Laboratory, University of Bristol, Bristol, United Kingdom\\
% ...
$^{57}$Department of Physics, University of Warwick, Coventry, United Kingdom\\
% ...
$^{60}$School of Physics and Astronomy, University of Glasgow, Glasgow, United Kingdom\\
$^{61}$Oliver Lodge Laboratory, University of Liverpool, Liverpool, United Kingdom\\
% ...
$^{69}$Syracuse University, Syracuse, NY, United States\\
% ...
$^{85}$Department of Physics and Astronomy, Uppsala University, Uppsala, Sweden, associated to $^{60}$\\
% ...
}
\end{flushleft}

\end{document}